\documentclass[prd,aps,preprintnumbers,floats,floatfix,superscriptaddress,preprintnumbers,showpacs,eqsecnum,nofootinbib,notitlepage,twocolumn]{revtex4} 

\usepackage[final]{graphicx}
\usepackage{dcolumn}
\usepackage{bm}
\usepackage{amsmath}	
\usepackage{amssymb}	
\usepackage{hyperref}
\usepackage{cleveref}
\usepackage{comment}
\usepackage{url}
\usepackage{float}
\newcommand{\beq}{\begin{equation}}
\newcommand{\eeq}{\end{equation}}
\usepackage{hyperref}
\usepackage{placeins}

\newcommand{\bea}{\begin{eqnarray}}
\newcommand{\ena}{\end{eqnarray}}
\newcommand{\beann}{\begin{eqnarray*}}
\newcommand{\enann}{\end{eqnarray*}}
\newcommand{\gsim}{\, \mbox{\raisebox{-1.ex}
{$\stackrel{\textstyle>}{\textstyle\sim}$}}\,}
\newcommand{\lsim}{\, \mbox{\raisebox{-1.ex}
{$\stackrel{\textstyle<}{\textstyle\sim}$}}\,}

\usepackage[dvipsnames]{xcolor}

\begin{document}


\title{Importance of tidal resonances in extreme-mass-ratio inspirals}

\author{Priti \sc{Gupta}} 
\email{priti.gupta@tap.scphys.kyoto-u.ac.jp}
\affiliation{
Department of Physics, Kyoto University, Kyoto 606-8502, Japan
}
\author{B\'eatrice \sc{Bonga}} 
\affiliation{
Institute for Mathematics, Astrophysics and Particle Physics, Radboud University, 6525 AJ Nijmegen, The Netherlands
}
\author{Alvin \sc{J. K. Chua}} 
\affiliation{
Theoretical Astrophysics Group, California Institute of Technology, Pasadena, CA 91125, United States
}
\author{Takahiro \sc{Tanaka}} 
\affiliation{
Department of Physics, Kyoto University, Kyoto 606-8502, Japan
}
\affiliation{Center for Gravitational Physics, Yukawa Institute for Theoretical Physics,
Kyoto University, Kyoto 606-8502, Japan
}

\date{\today}

\begin{abstract}
Extreme mass ratio inspirals (EMRIs) will be important sources for future space-based gravitational-wave detectors. In recent work, tidal resonances in binary orbital evolution induced by the tidal field of nearby stars or black holes have been identified as being potentially significant in the context of extreme mass-ratio inspirals. These resonances occur when the three orbital frequencies describing the orbit are commensurate. During the resonance, the orbital parameters of the small body experience a ‘jump’ leading to a shift in the phase of the gravitational waveform.  In this paper, we treat the tidal perturber as stationary and restricted to the equatorial plane, and present a first study of how common and important such resonances are over the entire orbital parameter space. We find that a large proportion of inspirals encounter a low-order resonance in the observationally important regime. While the ‘instantaneous’ effect of a tidal resonance is small, its effect on the accumulated phase of the gravitational waveform of an EMRI system can be significant due to its many cycles in band; we estimate that the effect is detectable for a significant fraction of sources. We also provide fitting formulae for the induced change in the constants of motion of the orbit due to the tidal resonance for several low-order resonances.
\end{abstract}

\maketitle


\section{INTRODUCTION}
\label{sec:1}
The three observation runs by gravitational-wave (GW) observatories LIGO and VIRGO have unveiled an exciting number of detections~\cite{Abbott_2020,Abbott_2021}, thereby allowing probes of binary dynamics in the strongly gravitating regime and discovering more about binary formation channels~\cite{ligo2021population,theligo2020tests}. By the early 2030s, the Laser Interferometer Space Antenna (LISA) and Taiji/TianQin will probe the cosmos at lower frequencies ($\sim$ mHz range)~\cite{amaroseoane2017laser,berry2019unique,Mei_2020}. One of the promising and exciting sources for these space gravitational wave antennae is inspirals of stellar-mass compact objects of mass $\mu \sim 1$ - $100M_\odot$ into supermassive black holes (SMBHs) of mass $M \sim 10^{5}$ - $10^{7} M_\odot$.

At leading order in mass ratio, the smaller body can be treated as a point-like particle moving along a geodesic orbit around the large black hole. At subsequent orders, a `self-force' arises from the small body’s interaction with its own gravitational perturbation that moves the orbit away from the geodesic of the Kerr spacetime~\cite{Mino_1997,Quinn_1997,Poisson_2011,Barack_2018}. The dissipative piece of the self-force is predominantly responsible for the inspiral, while the conservative piece shifts the orbital frequencies. A typical EMRI is expected to spend more than a year in observational band and undergoes $\sim 10^{5}$ orbital cycles around the central massive black hole, {\it i.e.}, about $10^{6}$ radians in gravitational-wave phase. GWs from such inspirals carry intricate details about the curvature of black holes, hence offering high precision tests of General Relativity (GR) in the extreme mass ratio limit.

There are two independent channels to form an EMRI. The ``traditional'' channel operates through scattering and capture processes. These can put stellar-mass objects in galactic nuclei close enough to the central massive BHs in galactic centers for the object to be gravitationally bound to the SMBH~\cite{Amaro_Seoane_2019,amaroseoane2020gravitational,Emami_2020,emami2020detectability,pan2021formation,pan2021formation2}. Recently, an alternative formation channel for EMRIs around accreting massive black holes has been proposed~\cite{pan2021formation,pan2021formation2} and is referred to as the wet formation channel. In this channel, stellar-mass black holes (and stars) on inclined orbits are captured by the accretion disk, and under the influence of density wave generation and head wind migrate towards the central SMBH~\cite{Kocsis_2011}. Despite the fact that roughly $1\%$ local galaxies and $10\%$ high-redshifted galaxies have active galactic nuclei ~\cite{Galametz_2009,Macuga_2019}, this wet EMRI formation channel is fairly efficient and expected to be equally important (if not more important) as the traditional channel. 
The two formation scenarios have distinct characteristics: EMRIs formed in the dry environment of traditional capture channels are expected to have higher eccentricities and higher inclinations than EMRIs formed in the wet environment of accretion disks when they enter the LISA band. For this reason, capture channels are particularly interesting for our study. The EMRI event rate depends on the population of compact objects, their stellar density profile around each SMBH, and also the mass and spin of SMBH.  All of these properties are highly uncertain, even for our galaxy. According to~\cite{Gair_2017}, the detection rate of EMRIs formed through the traditional formation channel by LISA is estimated to be from a few tens to a few thousand per year, if the detection threshold of SNR is 20. 

It is unlikely that all EMRIs can be treated as completely isolated for the duration in the LISA band. For instance, studies based on a Fokker-Planck simulation suggest that a population of 40$M_\odot$ BHs can be close to Sagittarius $\,$A$\!^\star$, with a median distance $\sim$ 5$\,$AU~\cite{Amaro_Seoane_2011,emami2020detectability,byh}. According to~\cite{Amaro_Seoane_2019,Gourgoulhon_2019}, brown dwarfs can be at an approximate distance of $\sim$ 30$\,$AU for Sgr$\,$A$\!^\star$. If this holds for even 10\% of EMRI events, the detection rate for the  observation of tidal resonances can be approximated to be a few $\rm yr^{-1}$ \cite{byh}. If an EMRI system is not isolated but is instead influenced by another astrophysical object, the tidal perturbation (even though relatively small to the background) can modify the orbital dynamics and GW radiation of the EMRI system resulting in phase variations in the gravitational waveform~\cite{byh,PhysRevD.96.083015}. For an EMRI formed in a wet environment, the active accretion disk itself can be treated as a tidal perturber. Also, in this scenario, dynamical friction caused by the disk interaction may leave imprints on GWs~\cite{Kocsis_2011,zwick2021improved}. Recently, there has also been work focusing on the ``dephasing'' of EMRI signal due to the dynamical friction caused by dark matter halos around  SMBH~\cite{Eda_2013,Eda_2015,Kavanagh_2020}. All these effects are likely to be detectable with future GW observatories. 

We focus on tidal resonances caused by the tidal field generated by close stars/BHs near the EMRI system~\cite{byh}. During most of the EMRI inspiral, the tidal field of nearby objects can be neglected. However, when the three fundamental orbital frequencies describing the orbit become commensurate, a tidal resonance occurs\footnote{Tidal resonances occur under more general conditions than self-force resonances, which require $n\omega_{r} +k\omega_{\theta}=0$. A tidal resonance occurs when the three orbital frequencies of the EMRI \emph{and} the three of the perturber are commensurate \cite{Yang:2019iqa}. However, in this paper we treat the perturber as static, hence its corresponding orbital frequencies do not play a role in resonance condition.}. As a result, the gravitational potential of the tidal perturber measurably changes the orbit of the small BH and thereby the gravitational radiation it emits. GWs undergoing such resonances will therefore encode information --- although limited --- about the environment of the galactic center, which is difficult to obtain from electromagnetic observations.

To prepare for the upcoming low-frequency stage of GWs, we need our waveform models to be very accurate because gravitational wave observations rely on matched filtering techniques that are extremely sensitive to the phase of the gravitational waves emitted by the system. Accurate waveform modeling is not only required to extract the signal, but also a prerequisite to parameter estimation. Since the phase is directly related to the orbital evolution, it is necessary to take the tidal fields into consideration. 

Using the two-timescale expansion~\cite{PhysRevD.78.064028}, the orbital phase can be expanded with respect to the mass ratio $\eta= {\mu}/{M}$ (considering a body of mass $\mu$ orbiting an SMBH of mass $M$) as
\beq
\psi = \frac{1}{\eta} \left(\psi^{(0)}+\eta^{1/2} \psi^{\rm (res)}+\eta \psi^{(1)} + O(\eta^{3/2})\right),
\eeq
where ${\psi^{(0)}}/{\eta}$ denotes the orbital phase determined by the averaged dissipative piece of the first order self-force whereas $\psi^{(1)}$ denotes the post-adiabatic order derived from the remaining oscillatory piece of the first order self-force and dissipative piece of the second order self-force. 
Corrections to the phase due to resonance scale as the square root of the inverse of mass ratio. These corrections thus become large over an EMRI inspiral, dominating over post-adiabatic effects. Significant efforts focusing on the computation of the self-force are made by the community to model EMRI waveforms~\cite{Fujita_2020,hughes2021adiabatic,Chua_2021}. While self-force calculations are tedious,  resonances (both self-force and tidal) will further complicate this enterprise~\cite{PhysRevD.89.084036,PhysRevD.94.124042,byh,PhysRevLett.114.081102}. Recent work has shown the impact of self-force resonances on parameter estimation, suggesting that parameter estimates of a resonant EMRI orbit are likely to be biased if resonances are not taken into account in waveform modeling~\cite{speri2021assessing}.

In this paper, we develop analytic and numerical tools to study tidal resonances with the aim of surveying the orbital parameter space and investigating how often tidal resonances occur in realistic inspirals. We compute the accumulation in phase after a tidal resonance has been encountered by an EMRI to understand their impact on waveforms. We investigate properties of tidal resonances such as the effect of spin of the central massive black hole, and the orbital parameters of the EMRI on the strength of each resonance and the resulting phase shift.

The outline of the paper is as follows. In Sec.~\ref{sec:2}, we recall basic properties and evolution equations for Kerr geodesic motion and introduce the concept of tidal resonances. In Sec.~\ref{sec:3}, we describe the analytic and numerical computations to obtain the inspiral and change in conserved quantities. In Sec.~\ref{sec:4}, we present our results and show the dependence of tidal resonances and accumulated phase shift on orbital parameters. We also compare the analytical estimate of jump with the numerical code by implementing the tidal effects and 5PN equations of motion using the forced osculating orbital elements method. We summarise the results in Sec.~\ref{sec:5}. In the appendix ~\ref{appex:A}, we discuss the suppression of certain tidal resonances and provide fitting formulae for different resonances, respectively. Throughout this paper, we use geometrical units with $c=G =1$ where $c$ is the speed of light and $G$ is the gravitational constant.

\section{\label{sec:2}Formulation}
In this section, we begin with an overview of Kerr geodesics and set up the notation and conventions that we use. Next, the tidal force is added in the evolution equations leading us to the tidal resonance condition. We also discuss the relevant time scales and our assumptions about the tidal perturber.

\subsection{Bound geodesics}
Since the discovery of Kerr Solution in 1963, the Kerr black hole has been extensively studied~\cite{PhysRevLett.11.237,Teukolsky_2015}. We begin by summarizing the generic geodesic motion in Kerr spacetime~\cite{1972ApJBardeen,Schmidt_2002,Mino_2003,Fujita1_2009}. Consider a point-like body of mass $\mu$ orbiting a Kerr black hole described by mass $M$ and spin parameter $a$. We use Boyer-Lindquist coordinates $\{r$,$\theta$,$\phi\}$ and Mino time $\lambda$ to describe the geodesic equations:

\begin{subequations}
	\begin{align}
	\bigg(\frac{dr}{d\lambda}\bigg)^{2}
            &= \big[E(r^{2}+a^{2}) - a L_{z}\big]^2\nonumber\\
        &\qquad    - \Delta \big[r^2+(L_{z} - a E)^2 +Q \big]
            \nonumber\\
             &\hskip 0.06cm\equiv R(r)\,, \label{eq:geo1}\\
    \bigg(\frac{d\theta}{d\lambda}\bigg)^{2}
            &= Q -  {\rm cot}^{2} \theta L_{z}^{2}-a^{2} {\rm cos}^{2}\theta (1 -E^2)
            \nonumber \\   
             &\hskip 0.06cm \equiv \Theta(\theta)\,, \label{eq:geo2}\\
    \frac{d\phi}{d\lambda}
            &=\Phi_{r}(r)+ \Phi_{\theta}({\rm cos}\, \theta)-a\,L_{z}\,, \label{eq:geo3}\\
 	\frac{dt}{d\lambda} 
 	        &=T_{r}(r)+ T_{\theta}({\rm cos}\, \theta)-a\,E\,, \label{eq:geo4}
	\end{align}
\end{subequations}
The quantities $E, L_{z}$, and $Q$ are the orbit’s energy (per unit $\mu$), axial angular momentum (per unit $\mu M$), and Carter constant (per unit $\mu^2 M^2$).
Here, $\Delta = r^2 - 2Mr + a^2$ and the Mino time parameter $\lambda$ is related to proper time $\tau$ by $d\lambda = d\tau/\Sigma$, where $\Sigma = r^2 + a^2 {\rm cos^2} \theta$. The explicit forms of functions in Eqs. (\ref{eq:geo3}) and (\ref{eq:geo4}) can be found in Fujita \& Hikida's paper, Ref~\cite{Fujita1_2009}.

By introducing $\lambda$ the radial and angular equations of motion are completely decoupled as can be seen in Eqs. (\ref{eq:geo1}) and (\ref{eq:geo2}). Therefore, for a bound orbit, radial motion $r(\lambda)$ and angular motion $\theta(\lambda)$ become periodic functions with Mino-time periods $\Lambda_r , \Lambda_{\theta} $ defined as~\cite{Fujita1_2009},
\beq
\Lambda_{r} = 2 \int_{r_\mathrm{p}}^{r_\mathrm{a}} \frac{dr}{\sqrt{R(r)}}\,, \hskip .5cm
\Lambda_{\theta} = 4 \int_{\theta_\mathrm{\rm min}}^{\pi/2} \frac{d\theta}{\sqrt{\Theta(\theta)}}\,,
\label{eq:Minoperiods}
\eeq
where $r_\mathrm{a}$, $r_\mathrm{p}$ are the values of $r$ at the apoapsis and periapsis respectively and $\theta_{\rm min}$ is the minimum value of $\theta$ (measured from the black hole’s spin axis). The motion in $t$ and $\phi$ can be written as a sum of three parts: a linear term with respect to $\lambda$, an oscillatory radial part with period $\Lambda_r$, and an oscillatory angular part with period $\Lambda_\theta$ as follows:
\begin{eqnarray}
        &&t(\lambda) = t_{0} + \Gamma_{t} \lambda + t^{(r)}_{\lambda} + t^{(\theta)}_{\lambda}\,,
        \label{eq:tlambda}\\
        &&\phi(\lambda) = \phi_{0} + \gamma_{\phi} \lambda + \phi^{(r)}_{\lambda} + \phi^{(\theta)}_{\lambda}\,.
        \label{eq:philambda}
\end{eqnarray}
    
In the above equations, $t_{0}$ and $\phi_{0}$ describe the initial conditions. The quantities $\Gamma_{t}$ and $\gamma_\phi$ describe the frequency of coordinate time and $\phi$ with respect to $\lambda$, respectively, which are given by~\cite{Fujita1_2009}
\begin{eqnarray}
        &&\Gamma_{t} =  \langle T_r(r)\rangle_\lambda +  \langle T_\theta(\rm{cos} \theta)\rangle_\lambda + \textit{a}\textit{L}_\textit{z}\,,
        \label{eq:Gammat}\\
        &&\gamma_{\phi} =  \langle \Phi_r(r)\rangle_\lambda +  \langle \Phi_\theta(\rm{cos} \theta)\rangle_\lambda -   \textit{a}\textit{E}\,,
        \label{eq:gammaphi}
\end{eqnarray}
where $\langle\dots \rangle_\lambda$ represents the time average over $\lambda$.

The associated frequencies with Mino-time periods are given by
\beq
\gamma_{r,\theta,\phi} = \frac{2 \pi}{\Lambda_{r,\theta,\phi}}\,.
\eeq

\begin{figure}[h]
		\includegraphics[width=8.5cm]{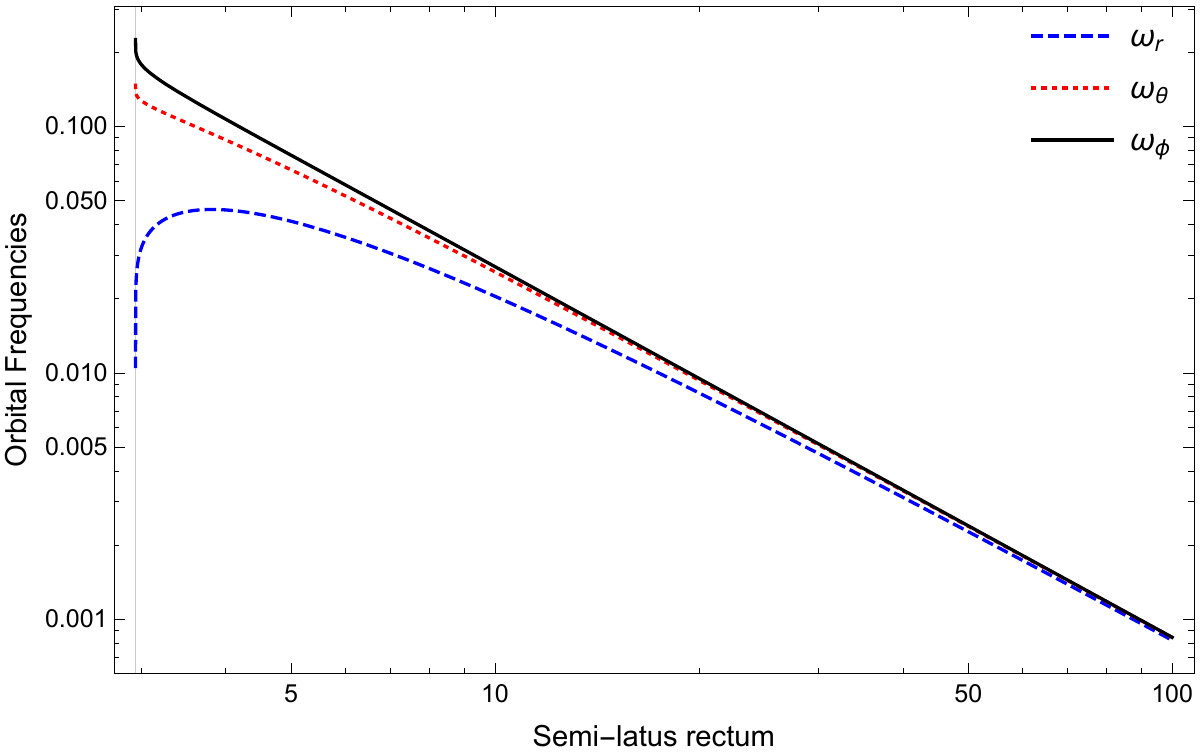}
\caption{Dimensionless fundamental frequencies as a function of semi-latus rectum for orbital eccentricity $e = 0.33$ and orbital inclination $30^{\circ}$. The spin parameter $a$ of central massive BH is set to be 0.9. The vertical grey line marks the location of the separatrix.}
		\label{fig:orbfreq}
	\end{figure}

The frequencies associated with distant observer time can be obtained by taking the ratio of the Mino-time frequencies to $\Gamma_{t}$:
\beq
\omega_{r,\theta,\phi} = \frac{\gamma_{r,\theta,\phi}}{\Gamma_{t}}\,.
\eeq
Unlike Keplerian orbits, bound Kerr geodesics are triperiodic. The radial frequency $\omega_r$ is associated with oscillations in the radial direction. The polar frequency $\omega_\theta$ is associated with oscillations in the $\theta$ direction, while the azimuthal frequency $\omega_\phi$ describes the rotations around the central BH spin axis. The frequencies of the precessional motions of the periastron and the orbital plane are $\omega_r-\omega_\phi$ and $\omega_\theta-\omega_\phi$, respectively. As shown in Fig.~\ref{fig:orbfreq}, in the weak field regime, these three frequencies asymptote to the frequency predicted by Kepler’s law whereas, in the strong field, they increasingly deviate from each other and evolve at different rates. Orbits are marginally stable at the separatrix and beyond this point, they become plunging orbits.

Besides the three constants of motion: $\{E,L_{z},Q\}$, the Kerr geodesic orbit can be characterised by another set of parameters: the semi-latus rectum $p$, the orbital eccentricity $e$, and orbital inclination angle $I$. These parameters are defined by
\begin{eqnarray}
       &&p := \frac{2 r_\mathrm{p}r_\mathrm{a}}{M (r_\mathrm{p}+r_\mathrm{a})}\,,\\
       &&e := \frac{r_\mathrm{a}-r_\mathrm{p}}{r_\mathrm{a}+r_\mathrm{p}}\,,\\
       &&I := \pi/2 - {\rm sgn}(L_z) \, \theta_{\rm min} \, .
\label{eq:pex}   
\end{eqnarray}
For later convenience, we also introduce  $x = \cos I$.

\subsection{Tidal resonances}
Gravitational waves from EMRIs will encode the information of curvature around the central black hole. In addition to this invaluable data, they can also be used to probe the stellar distribution in galactic centers. In our study, we consider an EMRI within the influence of an external tidal field. The information about the tidal environment created by a stellar-mass object near EMRI is treated in a fully relativistic framework by computing the perturbation to the Kerr spacetime (discussed in Sec.~\ref{sec:jump}).

The geodesic equations in Kerr are integrable, {\it i.e.}, there exists one integral of motion for each degree of freedom. The integrability allows one to introduce a set of “action-angle” variables, such that the “angle” variables $q_i$ parameterize a torus and the conjugate “action” variables $J_i$ are functions of the constants of motion $\{E, L_{z},Q\}$. This method is advantageous in obtaining the frequencies of Kerr orbits~\cite{Schmidt_2002} and including deviations to the geodesic motion due to different forces. Thus, we rewrite the EOM in this formalism to describe the dynamics in $(r,\theta,\phi)$~\cite{MTW_2017}.
\begin{eqnarray}
        &&\frac{dq_{i}}{d\tau}
            = \omega_{i} (\bold{J}) +\epsilon g_{i,\rm td}^{(1)}(q_\phi,q_\theta,q_r,\bold{J}) +
            \eta g_{i,\rm sf}^{(1)}(q_\theta,q_r,\bold{J})
         \nonumber \\   
         &&\hskip 0.7cm                                         
        + \hskip 0.1cm O(\eta^2,\epsilon^2,\eta\epsilon)\,,
\label{eq:EOM1}\\
        &&\frac{dJ_{i}}{d\tau}
            = \epsilon G_{i,\rm td}^{(1)}(q_\phi,q_\theta,q_r,\bold{J}) +
            \eta G_{i,\rm sf}^{(1)}(q_\theta,q_r,\bold{J})
         \nonumber \\   
         &&\hskip 0.7cm                                                         
        + \hskip 0.1cm O(\eta^2,\epsilon^2,\eta\epsilon)\,.
\label{eq:EOM2}
\end{eqnarray}
The parameter $\epsilon = M_{\star} M^2/R^3$ characterizes the strength of the tidal field produced by the perturber $M_{\star}$, and $R$ is the distance of the tidal perturber from $M$. As can be seen from the above equations, at zeroth order (on short timescales $\sim M$), a particle with mass $\mu$ is well approximated by a geodesic of the background spacetime. At this order, action variables are conserved, and $q_{i}$ increases at a fixed rate in time. However, in secular timescale ($\sim M/\eta$) the EMRI orbit deviates from geodesic motion due to the particle’s self-force ($g_{i,\rm sf}$,$G_{i,\rm sf}$)~\cite{Mino_1997,Quinn_1997,Poisson_2011,Barack_2018}. The leading order self-force motion is an adiabatic inspiral. Over the longer timescale, it is necessary to consider various post-adiabatic corrections currently under development~\cite{Pound_2020,upton2021secondorder}. Since we are interested in the tidal field from a nearby star or BH, another term denoting the tidal force is introduced in evolution equations ($g_{i,\rm td}$,$G_{i,\rm td}$). The tidal force depends on the axial position of the small body $\phi$ unlike the self-force (due to axisymmetry of the Kerr spacetime). The tidal force acts as a purely conservative force in contrast to the self-force which is both conservative and dissipative. Given the conservative nature of the tidal force, at leading order, the tidal force can be neglected throughout most of the inspiral except when a resonance is encountered (this is also  demonstrated in Fig.~\ref{fig:NumEvol}.) The mathematical description of the tidal resonance is similar to the resonance effect induced by the self-force itself~\cite{PhysRevLett.109.071102}. Both resonances are transient because the orbital frequencies are changing due to radiation reaction. The main difference between the two resonances is the force that causes it (the tidal force versus the self-force).

From here, we will focus on the tidal force $G_{i,\rm td}^{(1)}$ and drop the subscript `td'.  Each component of this force can be written as a Fourier series in terms of the angle variables
\beq
G_{i}^{(1)} (q_\phi,q_\theta,q_r,\bold{J}) = \sum_{m,k,n} G_{i,mkn}^{(1)}(\bold{J}) e^{i(mq_\phi + kq_\theta+nq_r)} \;. 
\label{eq:FT}
\eeq
For ergodic (non-resonant) orbits, the exponential term in the equation is rapidly oscillating in time averaging to zero over multiple orbits. Thus, generic $m,k,n$ modes do not contribute to secular change in $\bold{J}$. However, during an inspiral, it can happen that for a set of  integers $(m,k,n)$ 
\beq
\omega_{mkn} :=m \omega_{\phi}+k \omega_\theta +n \omega_r=0\,.
\label{eq:TR}
\eeq
 When this happens in the presence of a tidal perturber, a tidal resonance occurs. During resonance, the orbital motion is restricted to a subspace of the full orbital three-torus $\mathbb{T}^3=\{q_r,q_{\theta}, q_{\phi}\}$. When Eq.~\eqref{eq:TR} is satisfied, the phase in Eq.~\eqref{eq:FT} will be stationary near that time, and the exponential factor will vary slowly. The corresponding force amplitude $G_{i,mkn}^{(1)}$ is non-vanishing after averaging over many orbital cycles, and therefore induces a secular change in $\bold{J}$.
Generically for resonances, lower-order ones, {\it i.e.}, those with small integers $m$, $k$ and $n$ are more important than those with higher integers (this trend is also reported for self-force resonances~\cite{PhysRevD.94.124042} and mean-motion resonances~\cite{Yang:2019iqa}).

It is useful to mention the relevant timescales in our physical setting. The fastest timescales are the orbital periods $\sim \mathcal{O}(M)$ which can be defined using the three orbital frequencies as,
$$ T_r = 2\pi/\omega_r, T_\theta = 2\pi/\omega_\theta , T_\phi = 2\pi/\omega_\phi.$$
The radiation reaction (or slow) time $\tau_{rr}$ scales as $M/\eta$.
Another important time scale is the resonance duration $\tau_{\rm res}$. From the fact that the phase in Eq.~\eqref{eq:FT} changes slowly during a resonance, we can estimate its scale. In particular, expanding the phase variable $q_{mkn}:=m q_{\phi}+k q_\theta +n q_r$ in a Taylor series around the time at which the system encounters resonance, $\tau_{\rm res,0}$
\begin{align}
        &q_{mkn}(\tau) = q_{mkn}(\tau_{\rm res,0}) + (m\omega_{\phi}+k \omega_\theta +n \omega_r)(\tau-\tau_{\rm res,0}) \nonumber \\
        &\hskip 1.2cm+ \frac{1}{2}(m\dot{\omega}_{\phi}+k \dot{\omega}_\theta +n \dot{\omega}_r)(\tau-\tau_{\rm res,0})^2+\cdots\,.
        \label{eq:taylor}
\end{align}
The frequency and its derivative are evaluated at $\tau_{\rm res,0}$. For non-zero integers $m,k,n$, the second term $m\omega_{\phi}+k \omega_\theta +n \omega_r =0$ at  $\tau_{\rm res,0}$.
Thus, the duration of resonance is given by the condition that the third term becomes $\mathcal{O}(1)$, {\it i.e.},
\beq
\tau_{\rm res} \sim \sqrt{\frac{2}{m\dot{\omega}_{\phi}+k \dot{\omega}_\theta +n \dot{\omega}_r}} \sim M\sqrt{\frac{1}{\eta}}\, .
\eeq
Hence, the resonance time scale is longer than the orbital time scale and shorter than the radiation reaction time scale.
Lastly, another key timescale is the orbital period of tidal perturber $T_{\rm td} \sim 2\pi \sqrt{R^3/M}$. In our analysis, we ignore the dynamics of the tidal perturber. This assumption of a stationary third body is valid as long as $\tau_{\rm res}\gg T_{\rm td}$. However, if the third body is close to the EMRI on the equatorial plane, thereby violating the static approximation, the resonance condition is altered in the following way 
\beq
m (\omega_{\phi}\pm\Omega_{\phi,{\rm td}})+k \omega_\theta +n \omega_r=0 \, .
\eeq
In other words, the leading effect of the motion of the perturber would be the change in time of occurrence of resonance. Of course, the tidal force itself will also be different: instead of being time-independent, it will need to include the dynamical effects of the motion of the tidal perturber. However, the time-dependence of the tidal perturber 
is expected to be subdominant to the leading order quadrupolar field and therefore not considered in this paper (for a more extensive discussion about the modeling of the tidal field itself, see Sec.~\ref{sec:jump}).
Since for all resonances we consider $\Omega_{\phi,td} \ll \omega_{\phi}$, this shift is negligible in evaluating the resonance strength.
Note that the condition above is very similar to the resonance condition of mean motion resonances discussed in~\cite{Yang:2019iqa}. In fact, the tidal resonances considered in this paper are a subset of the relativistic mean motion resonances: tidal resonances are mean motion resonances for which the motion of the outer object can be considered static.

\section{\label{sec:3}Analytic and Numerical Implementation}
Here, we describe the methods used to model the tidal force and calculate the jump in conserved quantities due to a tidal resonance. We also discuss the procedure for determining EMRI inspiral orbits.

\subsection{The jump across tidal resonance}\label{sec:jump}
An EMRI can pass through a tidal resonance during the observationally relevant period. It can lead to a `jump' in constants of motion relative to the adiabatic prescription. After spending hundreds of orbital cycles in the resonance region, the parameters of the inspiraling orbit are different from those calculated from an adiabatic evolution. Flanagan and Hinderer~\cite{PhysRevLett.109.071102} gave an analytic expression for this deviation in the context of self-force resonances. The change across a tidal resonance is also well approximated by a very similar equation 
\begin{eqnarray}
\label{eq:Jump}
        &&\Delta J_{i}
            = \epsilon \int_{-\infty}^{\infty} G_{i}^{(1)} (q_\phi,q_\theta,q_r,\bold{J}) d\tau\nonumber\\
         &&\hskip 0.5cm                                                         
        = \epsilon \sum_{s=\pm 1} \sqrt{\frac{2 \pi}{|\Gamma s|}} {\rm exp}\bigg[{\rm sgn}(\Gamma s)\frac{i\pi}{4}+is\chi\bigg] \nonumber\\
        &&\qquad\qquad\qquad \times G_{i,sm,sk,sn}^{(1)}(\bold{J})\,.
\end{eqnarray}
Here, $\chi = mq_{\phi0} + kq_{\theta0}+nq_{r0} $ and $\Gamma = m\dot{\omega}_{\phi0} + k\dot{\omega}_{\theta0}+n\dot{\omega}_{r0}$, and the quantities $q_{i0}$ and $\dot{\omega}_{i0}$ are phases and frequency derivatives evaluated at $\tau_{\rm res,0}$ respectively. As discussed below Eq.~\eqref{eq:FT}, after long time averaging, only the components satisfying the tidal resonance condition contributes to a secular change in conserved quantities. Therefore, the jump across the resonance is evaluated by summing over non-vanishing harmonics of the tidal force $G_{i,mkn}$ after orbit averaging. In principle, $s$ ranges over all integers but since low-order resonances are dominant we only sum over $s=\pm 1$. All the quantities are evaluated at resonance. The change across resonance is proportional to $\epsilon/\eta^{1/2}$.

To calculate the tidal force $G_i^{(1)}$, we incorporate the influence of the third object, the tidal perturber, on the EMRI system by calculating its induced tidal deformation of the central BH spacetime. The induced deformation causes the small object of the EMRI to coherently accelerate when resonance occurs.
Thus as a first step, we need the perturbation $h_{\alpha \beta}$ to the central BH's spacetime due to the tidal field.  This is obtained by solving the Teukolsky equation~\cite{Teukolsky:1973} in the slow-motion limit (the radius of curvature $\mathcal{R}$ associated to the external spacetime is taken to be much larger than the BH’s scales, {\it i.e.}, $M/\mathcal{R} \ll 1$) followed by metric reconstruction so that the resulting metric is in the ingoing radiation gauge~\cite{Yunes2006}. Another metric describing a tidally deformed black hole given by Eric Poisson also exists~\cite{poisson2015tidal}, which is in the lightcone gauge with coordinates adapted to this gauge and does not rely on metric reconstruction. However, this metric is only valid in the slow spin limit and we would like to explore the entire range in spin of the central black hole. Therefore, we use the metric in~\cite{Yunes2006}\footnote{
Note that there is an overall factor of two missing in $h_{\alpha \beta}$ in~\cite{Yunes2006}; see footnote 17 in~\cite{LeTiec:2020bos} for details. After correcting for this factor, $dL_z/dt$ agrees in the slow spin limit with $dL_z/dt$ for $h_{\alpha \beta}$ given in~\cite{poisson2015tidal}.
}.

The metric given by~\cite{Yunes2006} includes only quadrupolar $l$=2 modes because the higher multipoles will be smaller by a relative factor of $\mathcal{O}(M/\mathcal{R})$. For $l$ =2, allowed values for azimuthal number $m$ are $-l$ to $l$. However, the $m$=0 mode is excluded from the metric\footnote{
These modes are included in the slow-spin limit metric given by Poisson~\cite{poisson2015tidal}.
}. For simplicity, we restrict the position of the tidal perturber to the equatorial plane. Under this restriction, the Newmann-Penrose scalar $\psi_0$ (see Eq.~(17) in~\cite{Yunes2006}) is zero for $m = \pm 1$ modes. As the metric does not contain $m = 0$ modes, we plan to include them in our future work. Therefore, the metric perturbation in our setting only contains $m = \pm 2$ modes.
 The input for the metric reconstruction procedure is $z_m$. At leading order, these coefficients are determined by the electric and magnetic quadrupole moment tensor denoted by $\mathcal{E}_{ab}$, and $\mathcal{B}_{ab}$, respectively (see Eq.~(7) in~\cite{Yunes2006}). Quadrupole moment tensors scale as~\cite{Poisson_2010}
$$\mathcal{E}_{ab} \sim \frac{1}{\mathcal{R}^2}\,,\hskip 0.5cm\mathcal{B}_{ab} \sim \frac{\mathcal{V}}{\mathcal{R}^2}\,,$$
where $\displaystyle\mathcal{V} \sim \sqrt{\frac{M+M_{\star}}{R}}$ is the orbital velocity of the third body. In this paper, we set the magnetic-type tensor to be zero as we assumed the tidal perturber to be stationary. In a general setting, the dynamics of the third body should be taken into account. To summarize our assumptions, we consider a stationary tidal perturber restricted to the equatorial plane and take into account only its $l=2$ and $m = \pm 2$  contributions in the tidal resonance. 

The tidal perturber is aligned along the $x$-axis and for the electric tidal moment tensor we take the following form:
\begin{align}
	\label{eq:tidal-tensor}
	\mathcal{E}_{ab}=  \frac{M_\star}{R^3} (2 \nabla_a x \nabla_b x - \nabla_a y \nabla_b y - \nabla_a z \nabla_b z)\,,
\end{align}
where $x$, $y$, and $z$ are the Cartesian-like coordinates (see Sec. IXB of \cite{Poisson2004}). We substitute this as input to obtain $h_{\alpha \beta}$ in the ingoing radiation gauge in advanced Eddington-Finkelstein coordinates (called Kerr coordinates in~\cite{Yunes2006}).

Next, we perform a coordinate transformation from the advanced Edington-Finkelstein coordinates $\{ v,r_{\rm EF},\theta_{\rm EF},\phi_{\rm EF} \}$ to Boyer-Lindquist coordinates $\{ t,r,\theta,\phi \}$:
\begin{subequations}
	\begin{align}
	dv &= dt + \left(1 + \frac{2 M r}{r^2 - 2 M r + a^2}\right) dr\,,\\
 	dr_{\rm EF} & = dr\,, \\
 	d\theta_{\rm EF} & = d\theta\,, \\
    d\phi_{\rm EF} & = d \phi + \frac{a}{r^2 - 2 M r + a^2} dr\,.
	\end{align}
\end{subequations}
Given $h_{\alpha \beta}$, the induced acceleration with respect to the background Kerr spacetime is
\begin{align}\label{eq:acc}
a^\alpha & = -\frac{1}{2} (g^{\alpha\beta}_{\rm Kerr}+u^\alpha u^\beta)(2h_{\beta \lambda;\rho}-h_{\lambda \rho; \beta}) u^\lambda u^\rho\;,
\end{align}
with $u^\alpha$ the unit vector tangent to the worldline of the EMRI's small mass $\mu$.
The instantaneous change rate of the constants of motion are~\cite{PhysRevD.96.083015}
\begin{align}
    \frac{dL_{z}}{d\tau}& = a_\phi\,,\label{eq:Ldot}\\
    \frac{dQ}{d\tau} &= 2 u_\theta a_\theta - 2 a^2 {\rm cos}^2 \theta u_t a_t + 2 {\rm cot}^2 \theta u_\phi a_\phi\,.
    \label{eq:Qdot}
\end{align}
The energy $E$ is conserved as the spacetime is stationary. With these equations in hand, we obtain $d{L}_z/d\tau$ and $dQ/d\tau$ due to the stationary phase harmonics of the tidal force, $G_{i,mkn}$, as a function of $\chi$ (see Eq.~\eqref{eq:Jump}).
Another quantity needed for the computation of jump  is $\Gamma$ which contains information about the resonance duration is obtained from the rate of change of the orbital frequencies at the time of resonance.

\subsection{Method of determining inspiral}
\label{subsec:MSTinsp}
For the evolution of an EMRI orbit, we use the numerical data for the
gravitational-wave fluxes dissipated by a stellar-mass object with bound orbits around a Kerr BH of spin parameter $a$ for large sets of orbital parameters. The derivation of GW fluxes in the data sets used methods presented in Refs.~\cite{Fujita_2004,Fujita_2005,Fujita_2009} based on the formalism developed by Mano, Suzuki, and Takasugi (MST)~\cite{Sasaki_2003,Mano_1996,Mano1_1996,Shuhei_1997}. The data shared with us by Fujita was produced for the extension of their recent paper dealing with equatorial inspirals at adiabatic order~\cite{Fujita_2020}.

Using the MST code, the adiabatic change of constants of motion was computed for a number of data points in the semi-latus rectum $p$, the orbital eccentricity $e$, and the orbital inclination $I$ for different spin parameters. We obtained $dC^{i}/dt$ in phase space $\{p,e, I\}$ through polynomial fitting where $C^{i}=\{E,L_z,Q\}$. Further, the secular evolution of orbital parameters $P^{i}=\{p,e,I\}$ is derived from those of $C^{i}=\{E,L_z,Q\}$ using
\beq
\bigg<\frac{dP^{i}}{dt}\bigg>= \sum_{j} (T^{-1})^i_{\,j} \bigg<\frac{dC^{j}}{dt}\bigg>,
\eeq
where $T^{\,j}_{i} \equiv \partial C^j/\partial P^j$ is the Jacobian matrix for the transformation from $\{E,L_z,Q\}$ to $\{p,e,I\}$. Using this approach, we obtain accurate orbits at inexpensive computational cost. One caveat is that the numerical data sets of GW fluxes are obtained only for orbital eccentricity $e$ upto 0.7 and each data set is truncated at $p \sim 6M$ for each value of the spin.\footnote{When numerical fluxes become available across the parameter space at inexpensive computational costs, we plan to extend our fittings in future work.} Therefore, accuracy of our numerical fitting for fluxes below $6M$ is limited by the available data sets and we rely on extrapolation for the change in fluxes in this region.

Returning to Eq.~\eqref{eq:Jump}, we obtain the change in frequencies during an inspiral from these numerical fits and evaluate $\Gamma$. For the implementation of the analytic expressions of fundamental frequencies~\cite{Fujita1_2009,Schmidt_2002}, our code employs the `Kerr Geodesic' Package from the Black Hole Perturbation Toolkit~\cite{BHPToolkit}. 
\begin{figure*}
\begin{center}
  \includegraphics[width=0.3\linewidth]{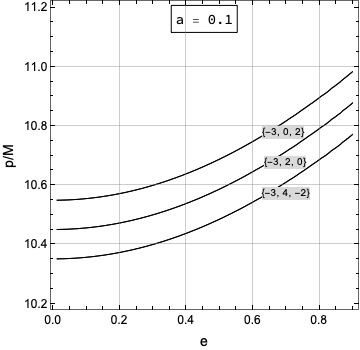}
    \hskip 0.5cm
\includegraphics[width=0.3\linewidth]{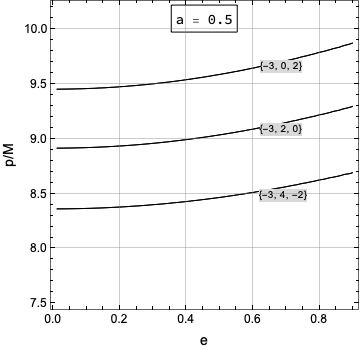}
 \hskip 0.5cm
\includegraphics[width=0.3\linewidth]{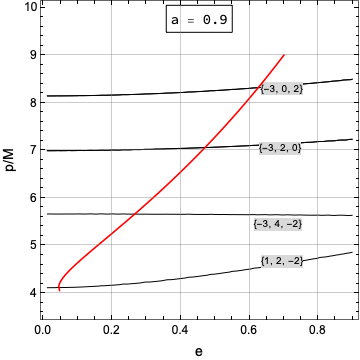}

\includegraphics[width=0.3\linewidth]{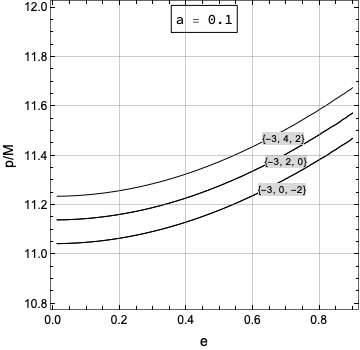}
    \hskip 0.5cm
\includegraphics[width=0.3\linewidth]{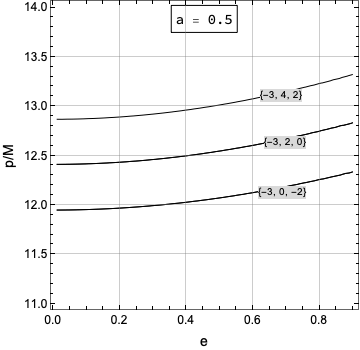}
 \hskip 0.5cm
\includegraphics[width=0.3\linewidth]{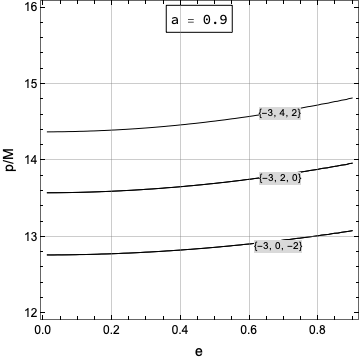}
 \caption{\small The upper panels shows the tidal resonance contours for prograde orbits with orbital inclination $50^{\circ}$ for different spin parameters of the central BH in $e$ - $p$ plane. The contours label correspond to integers $n,k,m$. In the right figure (upper panel), an inspiral is shown in red starting at $e=0.7$ and $p=9M$. During the evolution, $p$ shrinks and $e$ decreases due to radiation reaction. We see that before plunging, the orbit crosses multiple tidal resonances (We also show $m=0$ modes encountered by EMRIs. However, in our analysis we only consider tidal resonances with $m=\pm 2$ modes). The lower panels show the tidal resonance contours for retrograde orbits with orbital inclination $130^{\circ}$ for different spin parameters.}
\label{fig:TidalReso}
\end{center}
\end{figure*}

\begin{figure}
		\includegraphics[width=8.9cm]{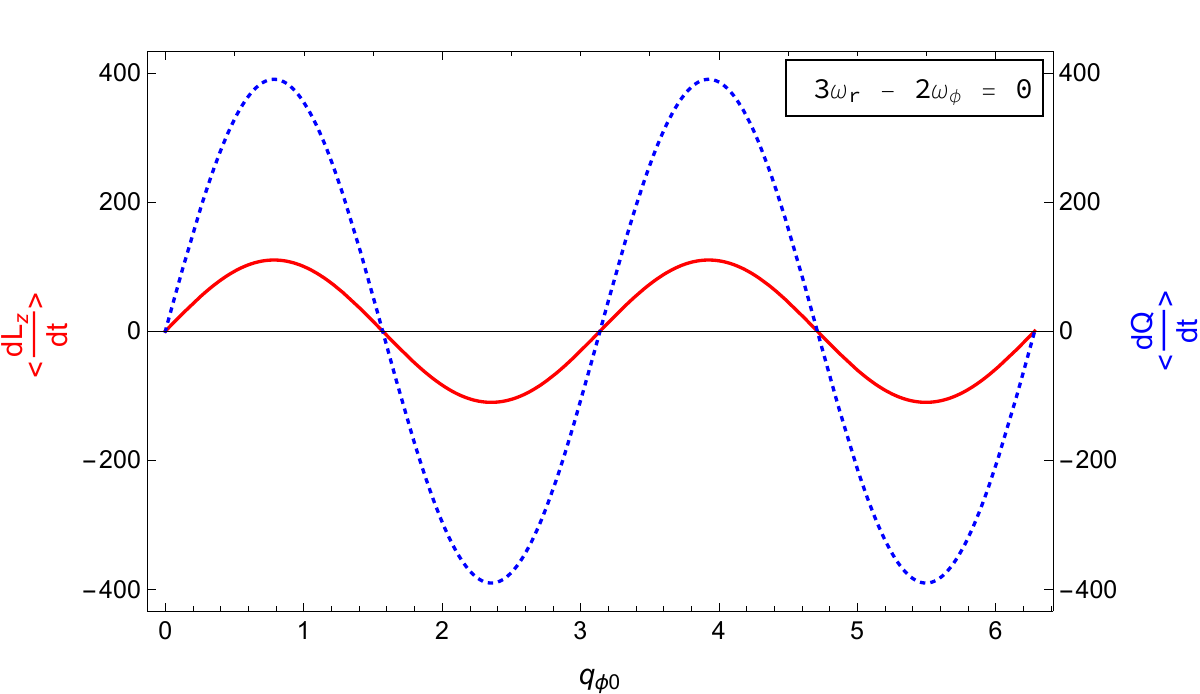}
\caption{\small Average change rate of z-component of angular momentum (red-solid) and Carter constant (blue-dotted) as a function of orbital phase $q_{\phi0}$ for an orbit crossing the $n:k:m =3:0:-2$ resonance with $a =0.9$. Both  $\langle dL_{z}/dt\rangle$ and $\langle dQ/dt \rangle$ are normalised by $\epsilon$ and powers of $M$ to be dimensionless.
}
		\label{fig:phasedep}
	\end{figure}
\section
{\label{sec:4}Results}
In this section, we investigate the orbital parameter space and find some   trends regarding the number of resonances encountered and the strength of each resonance as a function of the spin of the central massive black hole and the orbital parameters of the EMRI. We compute the accumulated phase shift due to different tidal resonances and show the affected parameter space. In addition to calculating the jump semi-analytically, we have also implemented the tidal effects using the forced osculating orbital elements method~\cite{osculating-schwarzschild,osculating-kerr}. The numerical evolution establishes that, as expected, the tidal force can be neglected throughout most of the EMRI evolution except during resonances. Moreover, the numerical evolution not only agrees qualitatively with the general features of tidal resonances, but also quantitatively. In particular, the numerically evaluated jumps agree remarkably well with the semi-analytic methods, thereby supporting the validity of both methods, which are implemented independently.

\begin{figure}
		\includegraphics[width=8.9cm]{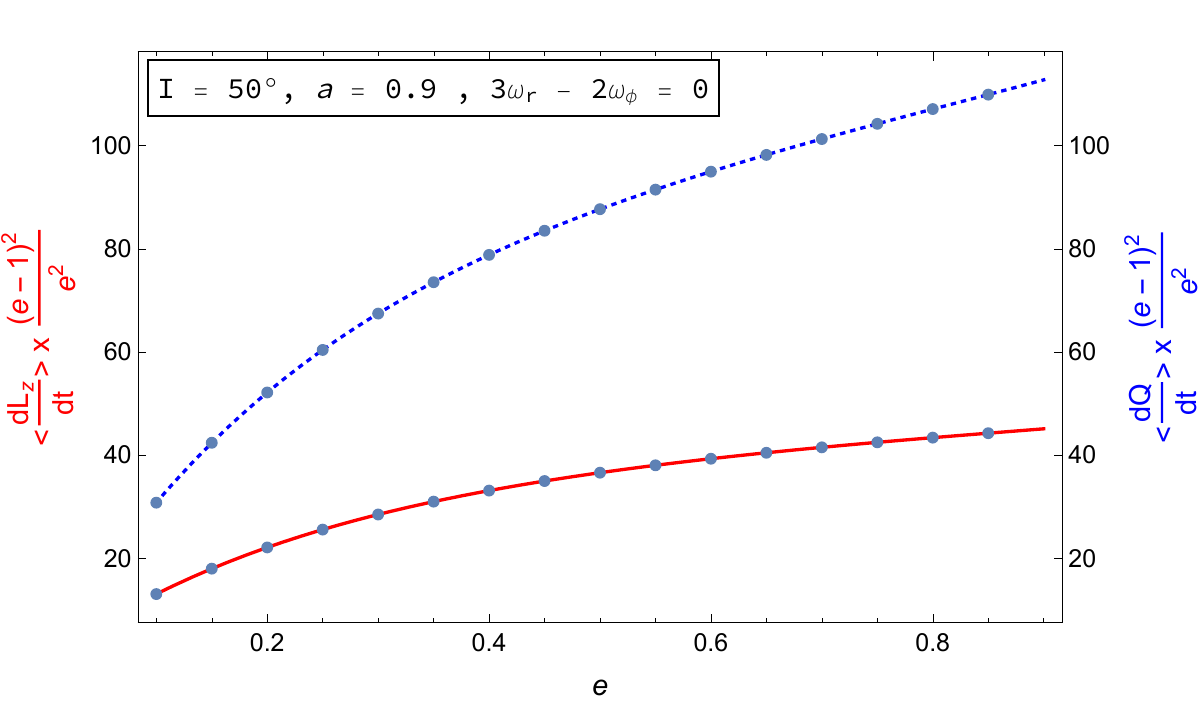}
\caption{\small Dependence of average change rate of the $z$-component of angular momentum (red-solid) and Carter constant (blue-dotted) on the orbital eccentricity for $n:k:m=3:0:-2$ with spin parameter and orbital inclination set to 0.9 and $50^\circ$, respectively. Both rates of change increase with increasing eccentricity. The factor $e^2/(e-1)^2$ ensures that $d{L}_{z}/dt$ and $dQ/dt$ are zero for circular orbits ($e=0$) since $\omega_r$ is zero in that case. The dots represent the values obtained from semi-analytic calculation and curves denote the obtained fitting.}
		\label{fig:LQvsEcc}
	\end{figure}
	
\begin{figure}
		\includegraphics[width=8.7cm]{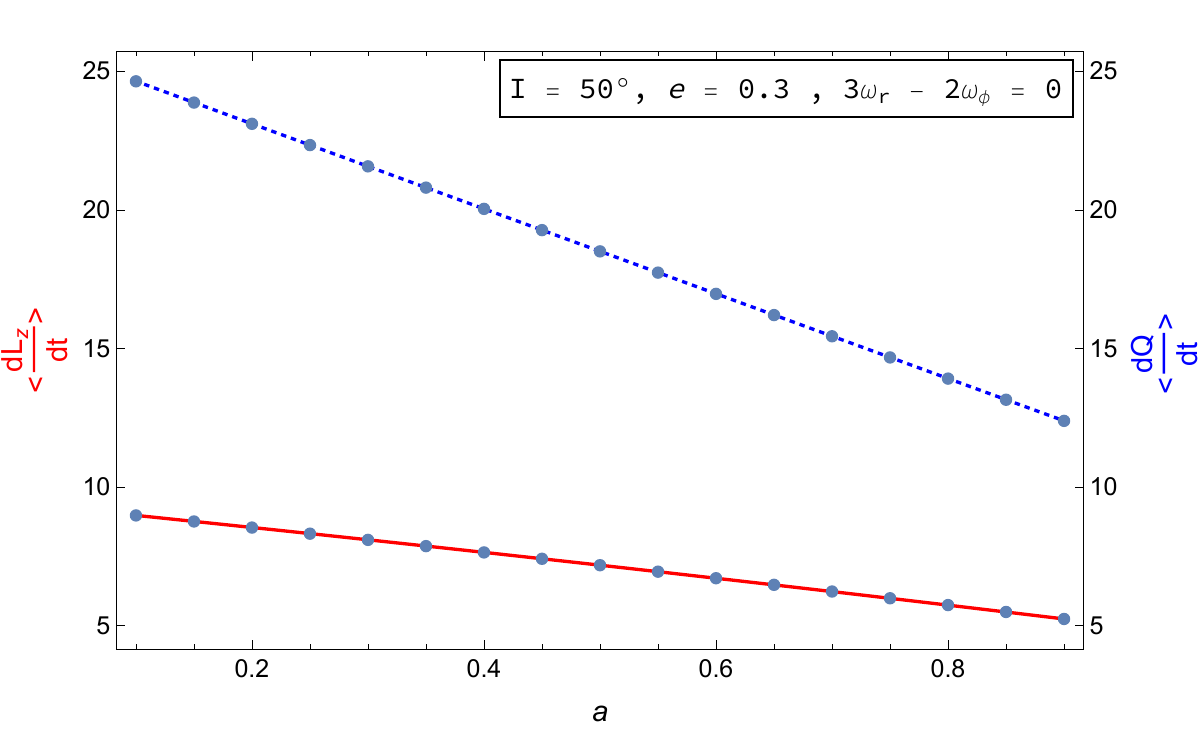}
\caption{\small Dependence of average change rate of the $z$-component of angular momentum (red-solid) and Carter constant (blue-dotted) on spin of central BH for  $n:k:m=3:0:-2$ with eccentricity and orbital inclination set to 0.3 and $50^\circ$, respectively. Both quantities decrease with increasing spin of SMBH.}
		\label{fig:LQvsSpin}
	\end{figure}
	
\begin{figure}
		\includegraphics[width=8.7cm]{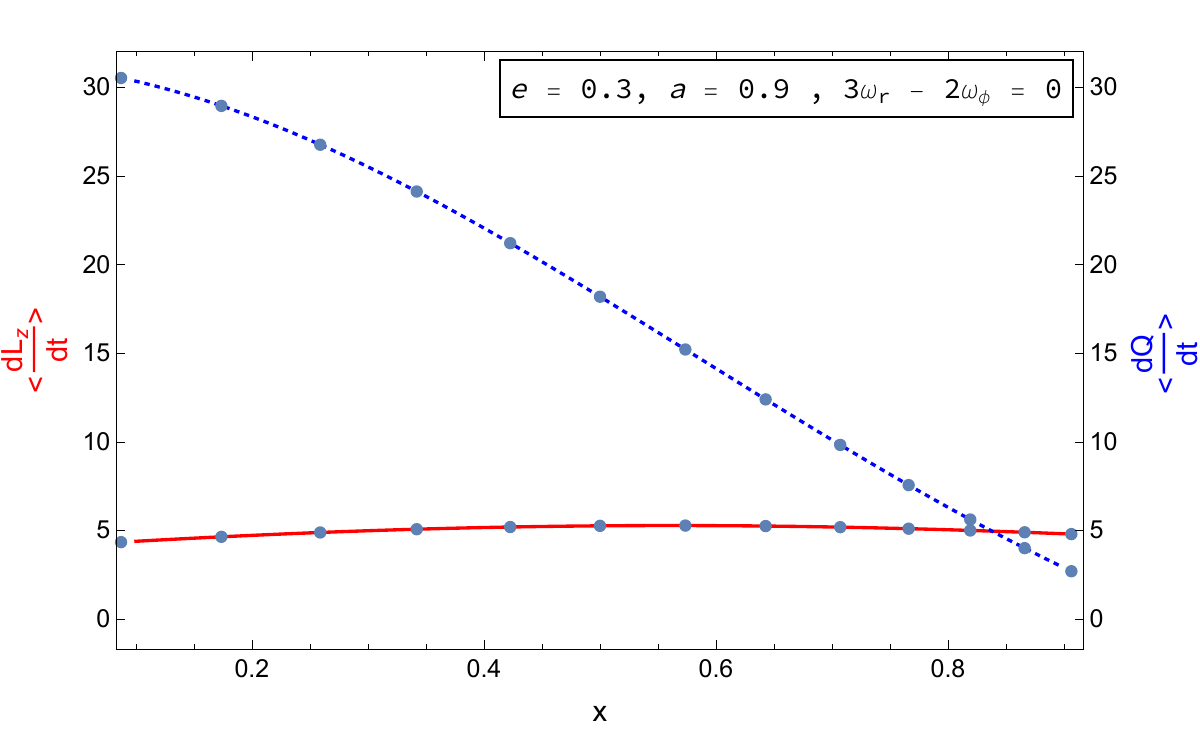}
\caption{\small Dependence of average change rate of the $z$-component of angular momentum (red-solid) and Carter constant (blue-dotted) on orbital inclination for  $n:k:m=3:0:-2$ with eccentricity and spin set to 0.3 and 0.9, respectively. As we go from high to a low inclination angle, $dQ/dt$ decreases whereas $d{L}_{z}/dt$ appears to be largely insensitive to the orbital inclination angle. The insensitivity of  $d{L}_{z}/dt$ to inclination angle is however only true for resonances with $k$ = 0.}
		\label{fig:LQvsInc}
	\end{figure}
	
\begin{figure}
		\includegraphics[width=8.7cm]{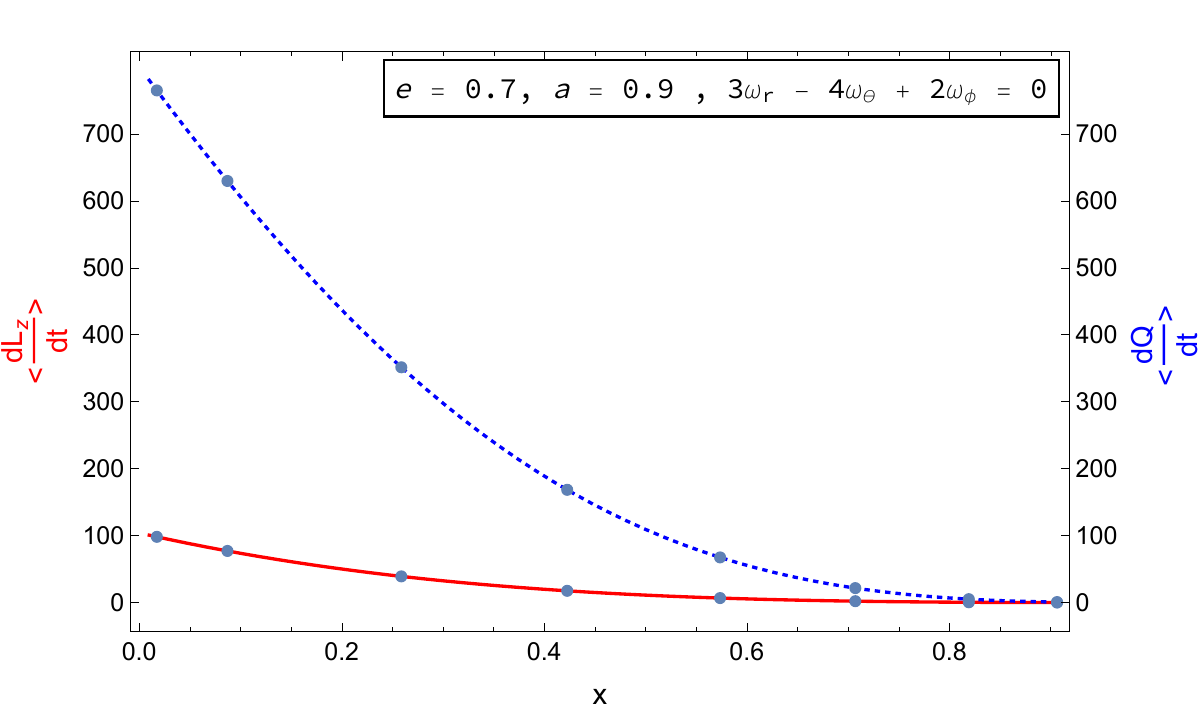}
\caption{\small Dependence of average change rate of the $z$-component of angular momentum (red-solid) and Carter constant (blue-dotted) on orbital inclination for  $n:k:m=3:-4:2$ with eccentricity and spin set to 0.7 and 0.9, respectively. As we go from high to a low inclination angle, $dQ/dt$ and $d{L}_{z}/dt$ decreases for the prograde orbit.}
		\label{fig:LQvsIncPro342}
	\end{figure}

\begin{figure}
		\includegraphics[width=8.7cm]{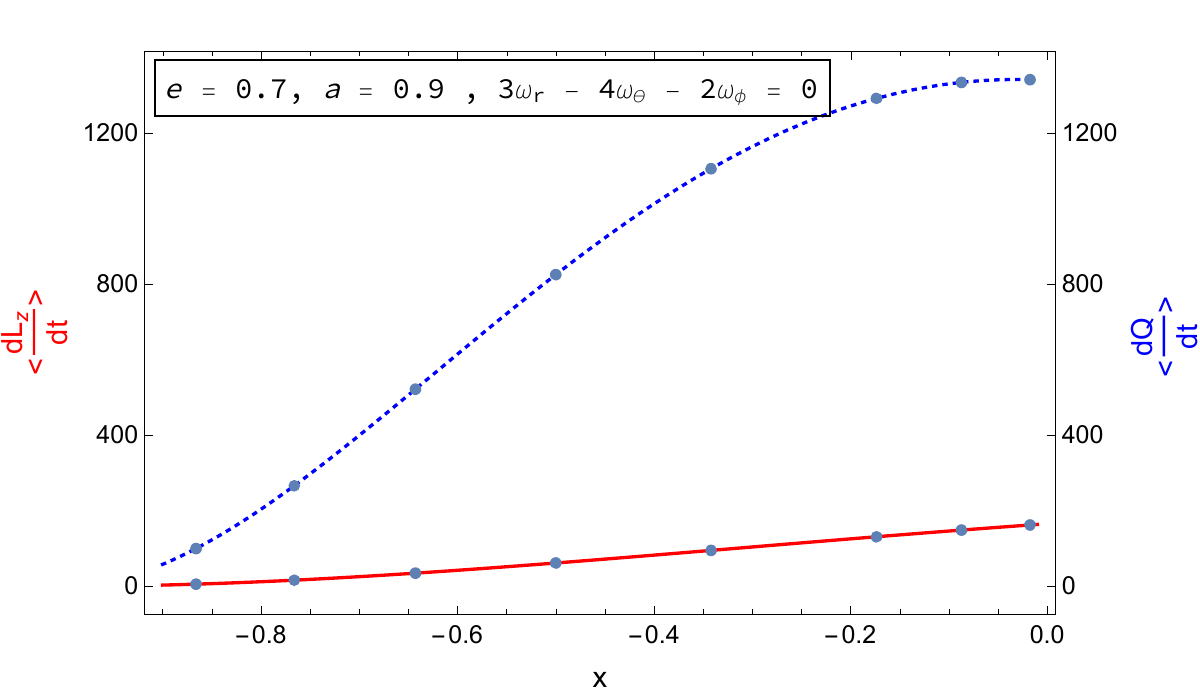}
\caption{\small Dependence of average change rate of the $z$-component of angular momentum (red-solid) and Carter constant (blue-dotted) on orbital inclination for  $n:k:m=3:-4:-2$ with eccentricity and spin set to 0.7 and 0.9, respectively. As we go from high to a low inclination angle, $dQ/dt$ and $d{L}_{z}/dt$ decreases for the retrograde orbit.}
		\label{fig:LQvsIncRet342}
	\end{figure}	

\subsection{Inspirals crossing tidal resonances}
Our aim is to span the complete orbital parameter space that is likely relevant for observationally important EMRI events. With the help of numerical data, we can compute the inspiral for both prograde ($0 \leq I < \pi/2$) and retrograde orbits ($\pi/2 < I < \pi$) by picking initial $I$ from the set $\in[20^\circ,50^\circ,80^\circ,100^\circ,130^\circ,160^\circ]$. The spin parameter are chosen from the set $a/M$ $\in [0.1,0.5,0.9]$ ranging from a slowly rotating central BH to a rapidly rotating one. For the orbital eccentricity $e$, the range varies from $0.0125\leq e \leq0.7$ with grid spacing $\Delta e = 0.0125$. The values of $p$ are not randomly sampled but are chosen such that the resonance condition in  Eq.~\eqref{eq:TR} is satisfied for some low order integers  $n,k$, and $m$.

We find that every inspiral encounters at least one lower-order resonance. As also seen for self-force resonances, higher-order resonances have smaller jumps compared to lower-order ones~\cite{PhysRevD.94.124042}. In Fig.~\ref{fig:TidalReso}, we show the low-order tidal resonances (i.e $n,k \in [-4,4]; m=0,\pm 2$) shown by black contours in the $e$ - $p$ plane for different spin parameters of the central black hole.  As discussed in Sec.~\ref{sec:jump}, when the tidal perturber is on the equatorial plane, $m=\pm 1$ modes are zero. In the upper panel, prograde geodesics are considered with $I = 50^\circ$ whereas in the lower panel, resonances are shown for retrograde geodesics with $I = 130^\circ$. We see that the value of $p$ at which resonances occur depends strongly on EMRIs orbital parameters. For instance, comparing the plots in the upper panel, the same resonance contour is in a different location on the $e$ - $p$ plane as the spin parameter varies (left to right) .

As an example, we show an inspiral (in red) evolving in the $e$ - $p$ plane with $a = 0.9$ and $I = 50^\circ$. As the orbit shrinks and circularizes due to radiation reaction it passes through four low-order tidal resonances before it plunges. When a resonance occurs at large $p$, the tidal field is stronger leading to a larger jump in conserved quantities. Note that for retrograde orbits (lower panel) resonances occur at larger values of $p$ as compared to prograde orbits thereby experiencing a larger tidal force. Also, at large $p$, the EMRI systems evolve relatively slowly, spending more time in resonance. To access the secular impact of tidal resonances on EMRIs the time remaining after crossing each resonance is also of importance. The space-based low-frequency interferometers will be able to track the evolution of EMRI waves for years. In the example shown, for an inspiral with parameters $M =4 \times 10^6  M_\odot$ and $\mu = 30 M_\odot$ the observational time after crossing the $n:k:m = -3:0:2$ resonance is about $10\, \rm yrs$ whereas the $-3:4:-2$ resonance is crossed $\sim 1.5\, \rm yrs$ before plunge.

\subsection{Dependence on orbital phase}
When we introduce the tidal perturber on the equatorial plane, the  spacetime describing the central black hole and the tidal perturber is no longer axisymmetric. As shown in Eq.~\eqref{eq:EOM2}, the tidal force depends on the axial position of the small body. Hence, the change in conserved quantities is sensitive to EMRI's orbital phase on entering the resonance. To illustrate this dependence, we first compute $d{L}_{z}/dt$ and $dQ/dt$ for some resonance with non-zero $m,k,n$. After orbit averaging, the sum in Eq.~\eqref{eq:FT} can be written as,
\begin{align}
&\big<G_{i}^{(1)} (q_\phi,q_\theta,q_r,\bold{J})\big> \cr 
 &\qquad \approx G_{i,mkn}^{(1)} (\bold{J})e^{ i(m q_{\phi0}+k q_{\theta0}+n q_{r0})} + \{\rm c.c.\}.
\end{align}
In Fig.~\ref{fig:phasedep}, we show dependence of average change rate of conserved quantities on $q_{\phi0}$ for an inspiral orbit (shown in Fig.~\ref{fig:TidalReso}) crossing the $3:0:-2$ resonance with $a =0.9$. Note that $\langle dL_{z}/dt\rangle$ and $\langle dQ/dt \rangle$ are made non-dimensional by factoring out $\epsilon/M$.

The phases $q_{r0}$ and $q_{\theta0}$ determine the values of $r$ and $\theta$ at resonance. Here, we set $q_{\theta 0}=0$ and $q_{r0}=0$ when the orbit enters resonance meaning that the orbit enters resonance at $\theta=I$ and $r=r_{\rm min}$. The azimuthal phase $q_{\phi0}$ describes the motion of a small object with mass $\mu$ around the central BH spin axis. The change induced in constants of motion has sinusoidal dependence on phase, {\it i.e.}, $\sin(m\,q_{\phi0})$.  Therefore, depending on this phase an orbit may cross the tidal resonance without ``feeling" its effect. In our analysis, to determine the impact of tidal resonances, we will fine-tune the phase value such that the change in ${L}_z$ and $Q$ due to resonance is maximum. In that sense, our results show the upper limit of influence caused by these resonances. The phase dependence is easily retrieved by multiplying the results here by $\sin(m q_{\phi0}+k q_{\theta0}+n q_{r0})$.

\subsection{Trends and fitting formulae}
In addition to the information of orbital phase, to estimate the jump in the constants of motion induced (see Eq.~\ref{eq:Jump}) by tidal resonances, we need the rate of change in orbital frequencies ($\Gamma$) and tidal force amplitude $G_{i,mkn}$. First, we survey the orbital parameter space and compute $d{L}_{z}/dt$ and $dQ/dt$ for different resonances to find some interesting trends. Using the numerical data obtained by evaluating the analytic expressions given in Eqs.~\eqref{eq:Ldot} and \eqref{eq:Qdot} we made 3-D $\{a,e,x\}$ fitting formulas by making a polynomial ansatz of the form $C_{ijk} a^i e^j x^k$ up to some order in $i, j, k$  and then fitting the numerical data points simultaneously to obtain the coefficients $C_{ijk}$. These numerical fits allow inexpensive calculations of $d{L}_{z}/dt$ and $dQ/dt$ due to a tidal resonance. 

We find that for all the resonances encountered by an inspiral before plunge the change in $L_{z}$ and $Q$ increases as we go from low to high eccentricity regardless of the rotation direction of the orbit, {\it i.e.}, prograde or retrograde. In Fig. \ref{fig:LQvsEcc}, we show an increase in both quantities with eccentricity for the $3:0:-2$ resonance. The dots represent the values obtained from the semi-analytic calculations and curves denote the obtained fitting. The agreement between the semi-analytic evaluation and fitting agrees remarkably well with the error always less than $1\%$. 

Another interesting pattern is observed with a variation in the spin parameter of SMBH. As shown in Fig. \ref{fig:LQvsSpin}, for prograde orbits, $d{L}_{z}/dt$ and $dQ/dt$ decrease as the spin parameter increases. This change directly translates to the kick induced during resonance implying that for rapidly spinning central massive objects the resonance strength is smaller. However for retrograde orbits, $d{L}_{z}/dt$ and $dQ/dt$ increase as the spin parameter increases. This is expected because the resonance occurs at larger $p$ values (see low panel Fig.~\ref{fig:TidalReso}). Thus, the acting tidal force is greater for retrograde orbits.

We also find that as the orbital inclination angle is varied from high to low, $dQ/dt$ and $d{L}_{z}/dt$ decreases for both prograde and retrograde orbits. In Fig \ref{fig:LQvsInc}, we show both the quantities for the $3:0:-2$ resonance. The change in ${L}_z$ appears to be insensitive to change in inclination, but it is only true for resonances with $k=0$. In Fig. \ref{fig:LQvsIncPro342}, dependence of $dQ/dt$ and $d{L}_{z}/dt$ on the orbital inclination is shown for the prograde orbit crossing $~3~:~-~4~:~2$ resonance. Similarly, the case for retrograde orbit  crossing $3:-4:-2$ resonance is shown in Fig. \ref{fig:LQvsIncRet342}. In our study, we found that resonances with $k$ = odd integers are suppressed, {\it i.e.}, they do not cause a jump in conserved quantities. This unique feature is discussed in Appendix \ref{appex:A}.

The fitting formulae to obtain change in $Q$ and ${L}_z$ by the $3:0:-2$ (prograde orbits) resonance are given by Eqs. \eqref{eq:FittingQ302Pro} and \eqref{eq:FittingLz302Pro} respectively. The fitting depends on orbital parameters $\{a,e,x\}$ and sinusoidally on orbital phases $q_{\phi0}$ and $q_{r0}$ at resonance. The prefactor $e^2/(e-1)^2$ ensures that $d{L}_{z}/dt$ and $dQ/dt$ are zero for circular orbits ($e=0$) since $\omega_r$ is zero for this case. Note that $\langle dL_{z}/dt\rangle$ and $\langle dQ/dt \rangle$ are normalised by multiplying a factor of $(\epsilon/M)^{-1}$. The Mathematica notebook with fitting formulae for other resonances (including $3:0:2$) is made available on \cite{BHPC}.
\subsection{Computation of induced jump and consistency with numerical evolution}\label{sec:num}
The estimate of induced jump in conserved quantities across a resonance is evaluated using the analytical expression given by Eq.~\eqref{eq:Jump}. For example, using this expression for an orbit crossing the $3:0:-2$ tidal resonance, the maximum jumps (by setting $q_{r0}=q_{\theta 0} =0, q_{\phi 0} \sim 0.785$) induced in $L_{z}$ and $Q$ are 
$$\Delta L_{z,max} = 7.4 \times 10^{-6} ,\hskip0.5cm \Delta Q_{\rm max} =1.8 \times 10^{-5}\,.$$
The above values are shown for an EMRI with mass ratio $\eta =7.5\times10^{-6}$ (for $M=4 \times 10^6 M_\odot \, \rm{and} \, \mu = 30 M_\odot$) and orbital parameters $\{a, p, e, x\}\sim\{0.9, 8.35, 0.62, 0.643\}$ at resonance under influence of a tidal perturber with mass $30 M_{\odot}$ at a distance of 10$\,$AU from the SMBH.

To perform a consistency check on the analytical calculation, we separately implemented the tidal force computed from the metric perturbation $h_{\alpha \beta}$ using the forced osculating orbital elements method~\cite{osculating-kerr,PhysRevD.77.044013}. For the inclusion of radiation reaction effects, we employ a newly developed solver of the PN fluxes that takes into account the correction up to 5PN order and tenth order in eccentricity~\cite{Fujita_2020,BHPC}. We use 5PN fluxes to drive the inspiral in our osculating code instead of MST fluxes because PN fluxes are easier to implement and MST flux data sets are limited to $p \sim 6M$. In the osculating geodesics approach, the instantaneous tangential geodesics are referred to as osculating orbits. The transition between osculating orbits corresponds to the change in orbital elements. The inspiral motion is constructed from a smooth sequence of tangent geodesics where the driving forces are radiation reaction (5PN fluxes) and the tidal force caused by the perturber. We ran two simulations for an inspiral orbit with and without the effect of the tidal force taking the same initial conditions for the orbit as shown in Fig.~\ref{fig:TidalReso}.  To extract the size of the jump, we compute the difference between the full trajectory (tidal force + 5PN) and adiabatic (only 5PN) trajectory.

\begin{figure*}
  \centering
  \includegraphics[width=0.47\linewidth]{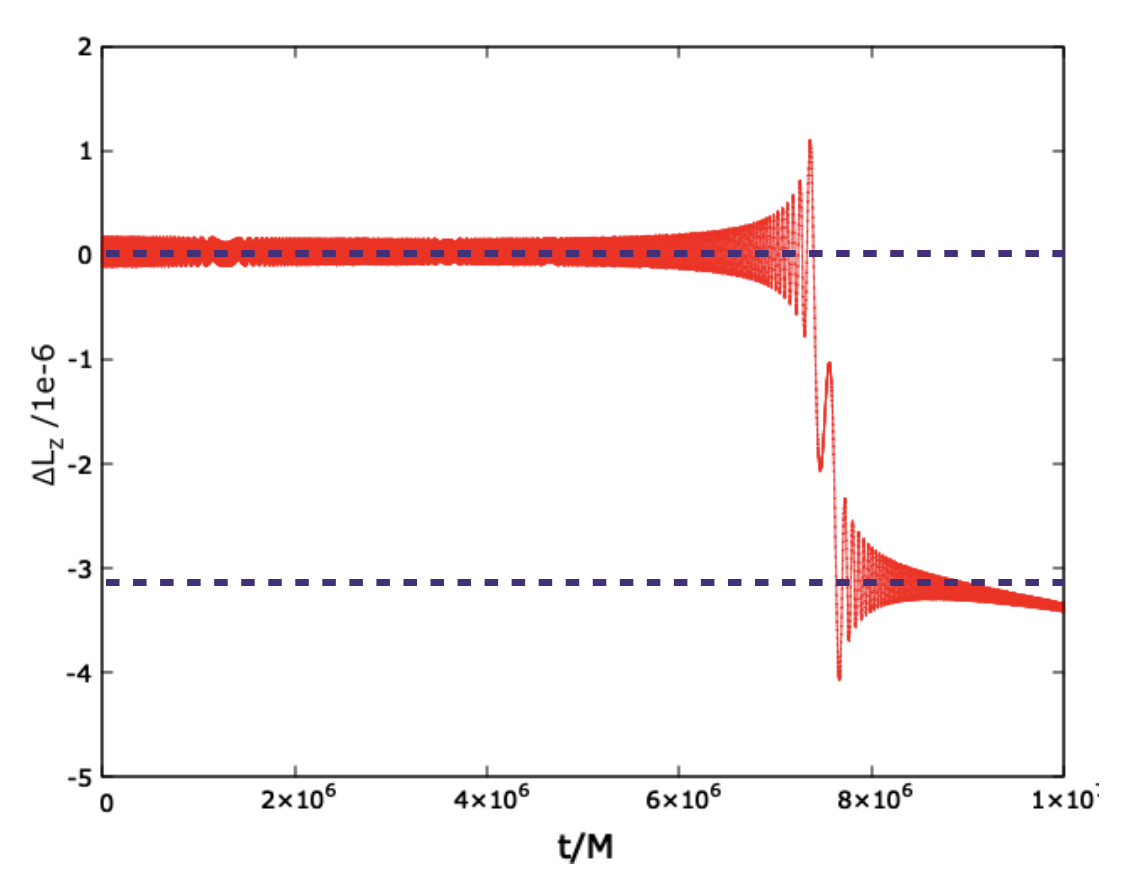}
    \hskip 0.5cm
\includegraphics[width=0.47\linewidth]{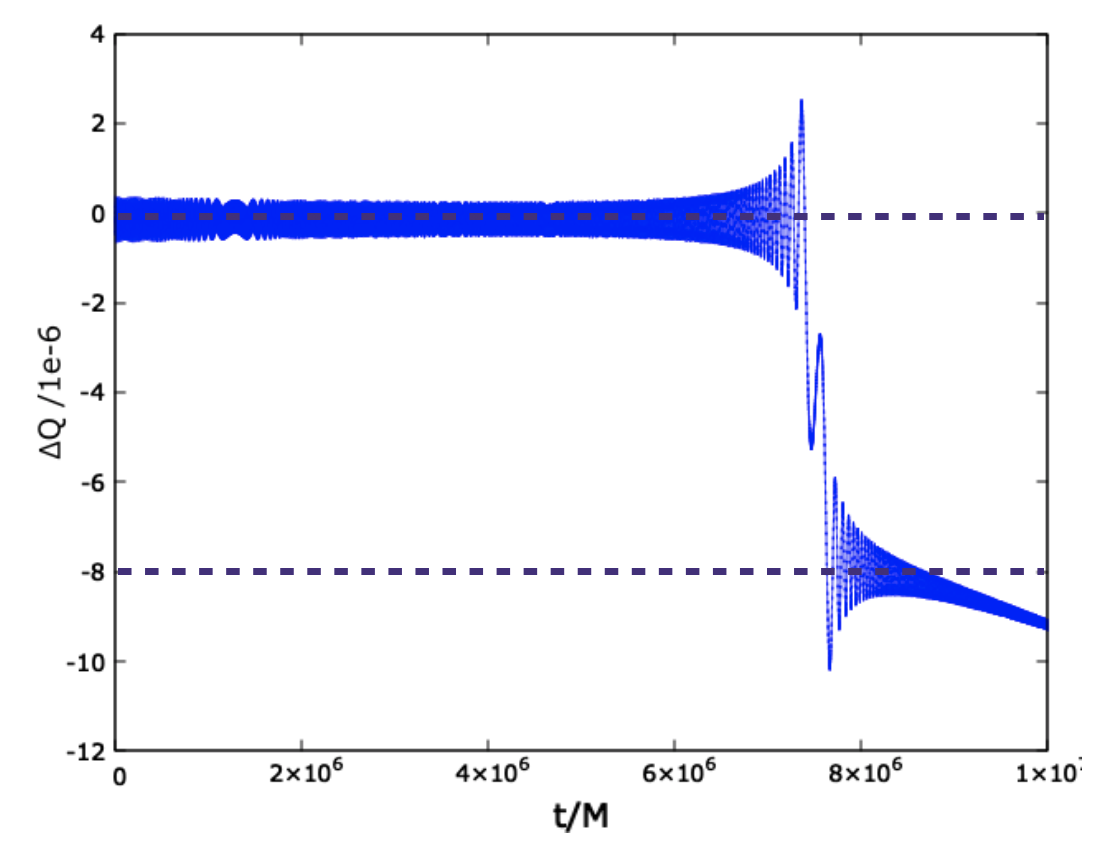}
\caption{\small The left figure shows the difference in $L_{z}$ between the orbit evolved with and without tidal resonance effect. When the orbit enters resonance, there is a jump in the quantity. The fast oscillations correspond to orbital timescales. The gap between the horizontal dotted lines estimates the size of the jump. Similarly, the right figure shows a jump in the Carter constant.}
\label{fig:NumEvol}
\end{figure*}

In Fig.~\ref{fig:NumEvol}, we show the differences $\Delta L_z$ (left) and $\Delta Q$ (right). The apparent thickness of the lines shown in the figures is caused by oscillations on the orbital timescale. The orbit spends hundreds of cycles in the resonance regime which lasts about 17 days. It also shows that the tidal force significantly affects the inspiral around the resonance only.

An EMRI orbit can enter the resonance with any orbital phase thus affecting the size of the jump. We first find the value of $q_{\phi0}$ at which $\Delta Q_{\rm max}$ matches $\Delta Q$ in the plot (right panel of Fig.~\ref{fig:NumEvol}) by solving $$\Delta Q_{\rm max} \sin(-2 q_{\phi0})=\Delta Q.$$ 
This yields $q_{\phi0} \sim 0.23.$ Then, we use this phase to check what the numerical value of $\Delta L_z$ should be based on the maximum value it can take analytically, {\it i.e.}, $\Delta L_{z,max}$. Our check yields $\Delta L_{z} \sim 3.2 \times 10^{-6}$, which agrees with the jump estimated from numerical evolution (left panel of Fig.~\ref{fig:NumEvol}). This computation verifies the jump estimated using the semi-analytic expression. Hereafter, we rely on the semi-analytical estimate of the jump (obtained using the numerical strategy discussed in Sec. \ref{subsec:MSTinsp}) to study the impact of tidal resonances on gravitational waves. However, the numerical osculating code is being used in our ongoing work to perform a more detailed investigation of strategies and implications for the modeling and analysis of tidally perturbed EMRIs \cite{PBAT} (see Sec. \ref{sec:5}).
\begin{widetext}
\begin{align}
    \begin{split}
    \left<\frac{dQ}{dt}\right> &=  \frac{e^2}{(e-1)^2} \bigg(6166.4  (a^2 (e^5 (1. x^6-3.3657 x^5+4.2989 x^4-2.64672 x^3+0.82724 x^2-0.11322 x-0.0325)\\
    &+e^4 (-2.561x^6+8.6411 x^5-10.9856 x^4+6.73865 x^3-2.14181 x^2+0.2942 x+0.01469)+e^3 (2.4592 x^6\\
    &-8.338 x^5+10.4957 x^4-6.38662 x^3+2.10369x^2-0.307742 x-0.02541)+e^2 (-1.0576 x^6+3.73285 x^5\\
    &-4.57117 x^4+2.7179 x^3-0.98419 x^2+0.16744 x+0.023683)+e (0.2142x^6-0.77253 x^5+0.858453 x^4\\
    &-0.46584 x^3+0.23502 x^2-0.054646 x-0.0146077)-0.0131998 x^6+0.0442585 x^5-0.057625 x^4\\
    &+0.0360626 x^3-0.01073x^2+0.0130161 x-0.00683955)+a (e^5 (-1.41091 x^6+4.70494 x^5-5.98812 x^4\\
    &+3.7815 x^3-1.17999 x^2+0.082132 x+0.01024)+e^4(3.63694 x^6-12.1218 x^5+15.3555 x^4-9.7812 x^3\\
    &+3.11081 x^2-0.16016 x-0.03945)+e^3 (-3.51743 x^6+11.7153 x^5-14.6986 x^4+9.5462x^3-3.1598x^2\\
    &+0.048679 x+0.06504)+e^2 (1.56518 x^6-5.20606 x^5+6.36726 x^4-4.35903 x^3+1.5984 x^2+0.10196 x\\
    &-0.06398)+e(-0.31407 x^6+1.02894 x^5-1.14294 x^4+0.9495 x^3-0.45183 x^2-0.10985 x+0.03645)\\
    &+0.02657 x^6-0.066807 x^5+0.0859032 x^4-0.0523576
   x^3+0.0151582 x^2-0.00205405 x+0.0944331)\\
   &+e^5 (0.50946 x^6-1.6744 x^5+2.1131 x^4-1.3159 x^3+0.3583 x^2+0.00554x+0.04718)+e^4 (-1.32659 x^6\\
   &+4.35489 x^5-5.48625 x^4+3.45939 x^3-0.77303 x^2-0.060315 x-0.168315)+e^3 (1.3068 x^6-4.2645x^5\\
   &+5.36471 x^4-3.47472 x^3+0.6569 x^2+0.15588 x+0.26122)+e^2 (-0.58987 x^6+1.944 x^5-2.42323 x^4\\
   &+1.67565 x^3-0.186175 x^2-0.1797x-0.22711)+e (0.12024 x^6-0.39217 x^5+0.49435 x^4-0.42018 x^3\\
   &-0.0448911 x^2+0.118569 x+0.12802)-0.00827 x^6+0.02662 x^5-0.03651 x^4+0.0195 x^3-0.05746 x^2\\
   &+0.000836659 x+0.000193748)\bigg) \sin(-2 q_{\text{$\phi $0}}+3q_{\text{$r$0}})\,,
   \end{split}
   \label{eq:FittingQ302Pro}
\end{align}

\begin{align}
    \begin{split}
     \left<\frac{d{L}_{z}}{dt}\right> &=  \frac{e^2}{(e-1)^2}\bigg(13.8664 (a^2 (e^5 (x^4+1.70942 x^3-0.812785 x^2-0.538936 x-0.32076)+e^4 (-4.11606 x^4\\
     &-4.79651 x^3+1.781 x^2+2.27585x+1.11764)+e^3 (7.17415 x^4+5.67992 x^3-1.22333 x^2-4.01184 x\\
     &-1.64238)+e^2 (-7.25395 x^4-3.92149 x^3-0.0528632 x^2+3.96175
   x+1.36426)+e (4.21764 x^4\\
   &+1.69592 x^3+0.404326 x^2-2.33394 x-0.756895)+0.012012 x^4-0.0125696 x^3+0.0127552 x^2\\
   &-0.00766985x-0.000627702)+a (e^5 (0.289607 x^4-3.94961 x^3-3.9027 x^2-1.82132 x+0.710913)\\
   &+e^4 (-0.370237 x^4+12.6334 x^3+15.3809x^2+6.2625 x-2.47674)+e^3 (-0.377266 x^4-17.4934 x^3\\
   &-25.9795 x^2-9.31769 x+3.78052)+e^2 (1.09716 x^4+14.3769 x^3+25.2598
   x^2+7.96198 x-3.5384)\\
   &+e (-0.864082 x^4-7.40462 x^3-14.5161 x^2-4.3349 x+1.92547)-0.0109531 x^4+0.0114339 x^3\\
   &-0.0375848 x^2-0.00156484x+0.00169788)+e^5 (-0.328544 x^4+0.766588 x^3+1.98025 x^2+5.37844 x\\
   &+2.57726)+e^4 (0.850133 x^4-2.02349 x^3-7.73277 x^2-19.233
   x-9.3297)+e^3 (-0.8120 x^4+2.0503 x^3\\
   &+13.0406 x^2+29.8651 x+14.6449)+e^2 (0.320494 x^4-0.886721 x^3-12.3364 x^2-26.4222x-13.0807)\\
   &+e (-0.0228461 x^4+0.142487 x^3+7.20898 x^2+14.7695 x+7.35985)+0.00829816 x^4-0.0186416 x^3\\
   &+0.0256575 x^2+0.0190106
   x+0.0117907)\bigg) \sin(-2 q_{\text{$\phi $0}}+3q_{\text{$r$0}})\,,
   \end{split}
   \label{eq:FittingLz302Pro}
\end{align}
\end{widetext}

\subsection
{\label{subsec:4} Impact on gravitational waveform}
\begin{figure*}
  \centering
  \includegraphics[width=0.3\linewidth]{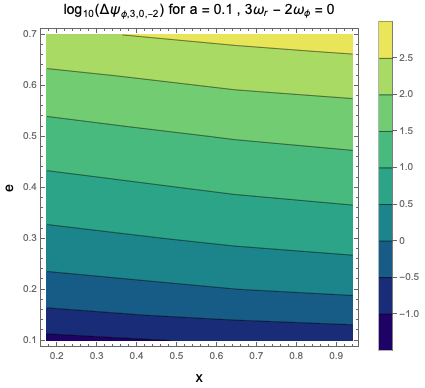}
    \hskip 0.5cm
\includegraphics[width=0.3\linewidth]{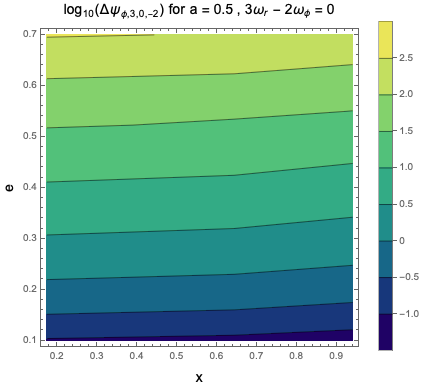}
\hskip 0.5cm
\includegraphics[width=0.3\linewidth]{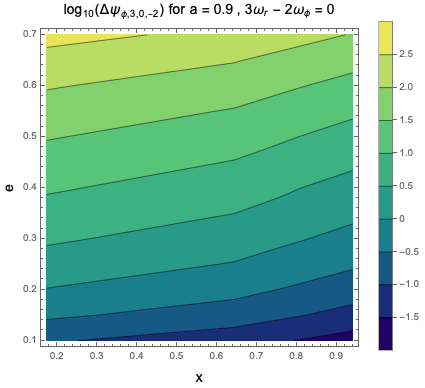}

\includegraphics[width=0.3\linewidth]{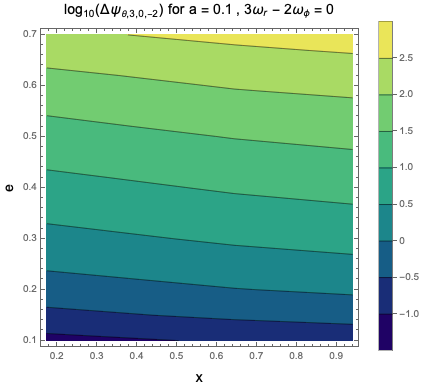}
    \hskip 0.5cm
\includegraphics[width=0.3\linewidth]{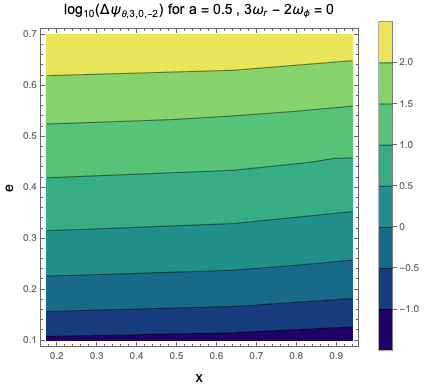}
\hskip 0.5cm
\includegraphics[width=0.3\linewidth]{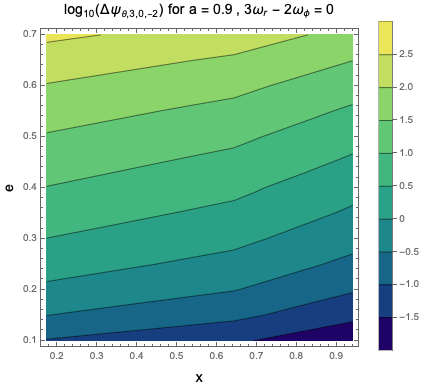}

\includegraphics[width=0.3\linewidth]{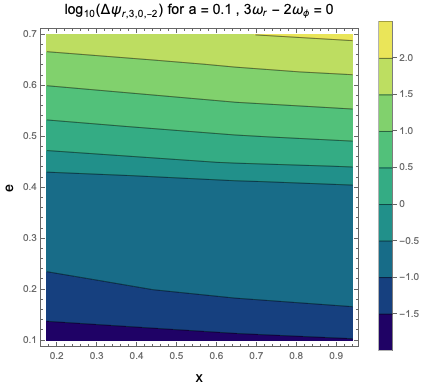}
    \hskip 0.5cm
\includegraphics[width=0.3\linewidth]{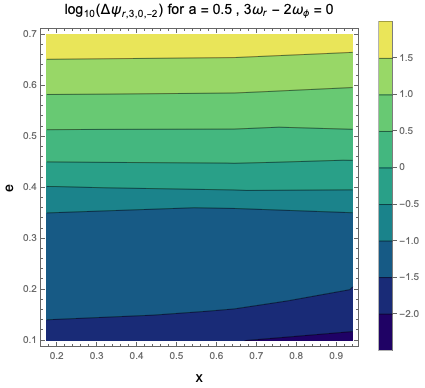}
\hskip 0.5cm
\includegraphics[width=0.3\linewidth]{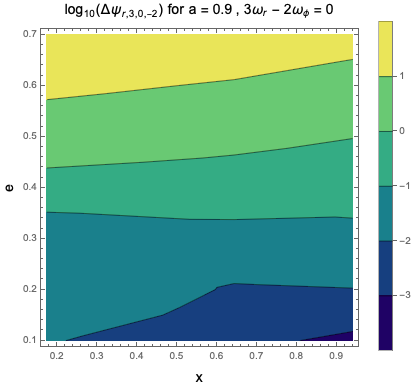}
\caption{Accumulated phase $\Delta \Psi_{i}$ for spin parameter $a = 0.1, 0.5, 0.9$ for a prograde orbit crossing the $3:0:-2$ resonance in the $x$ - $e$ plane. Top, middle and bottom panels correspond to $\Delta \Psi_{\phi}$, $\Delta \Psi_{\theta}$ and, $\Delta \Psi_{r}$, respectively. The phase shift is computed for an EMRI with $M=4 \times 10^6 M_\odot, \mu =30 M_\odot$ under the influence of a tidal perturber with mass $M_\star=30 M_\odot$ at a distance of 10$\,$AU from the central SMBH. Results for different sets of parameters can be estimated from the scaling relation given in Eq.~\eqref{eq:scale}.}
\label{fig:PSpro302}
\end{figure*}

\begin{figure*}
  \centering
  \includegraphics[width=0.3\linewidth]{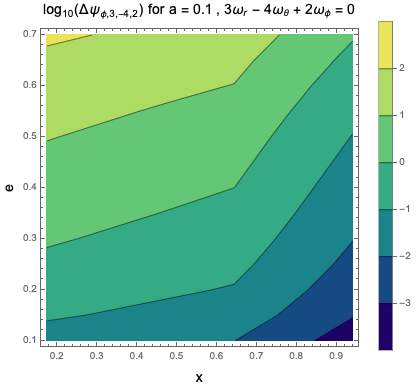}
    \hskip 0.5cm
\includegraphics[width=0.3\linewidth]{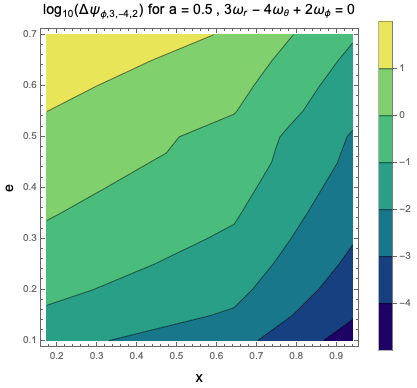}
\hskip 0.5cm
\includegraphics[width=0.3\linewidth]{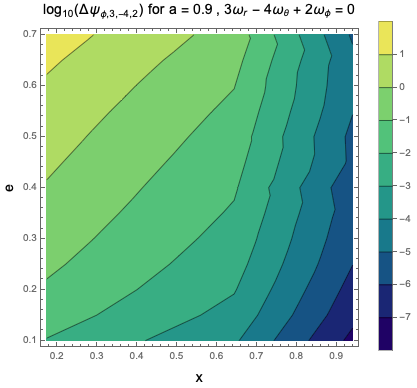}

 \includegraphics[width=0.3\linewidth]{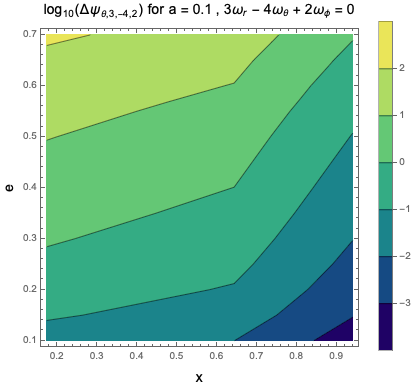}
    \hskip 0.5cm
\includegraphics[width=0.3\linewidth]{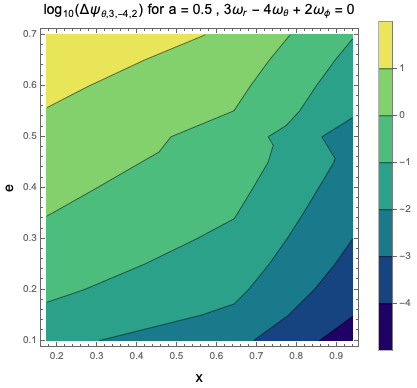}
\hskip 0.5cm
\includegraphics[width=0.3\linewidth]{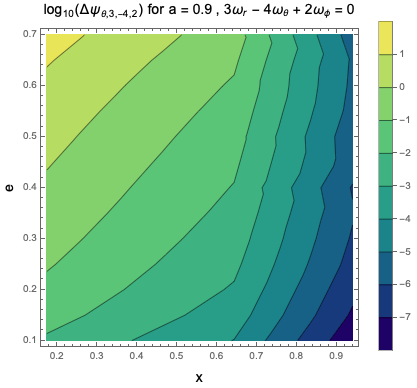}

 \includegraphics[width=0.3\linewidth]{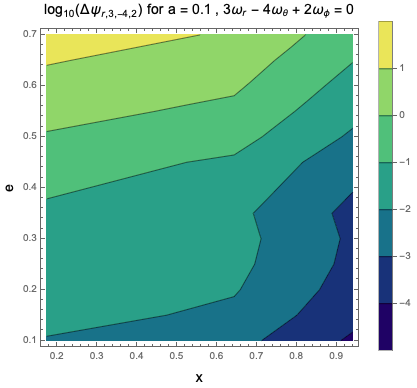}
    \hskip 0.5cm
\includegraphics[width=0.3\linewidth]{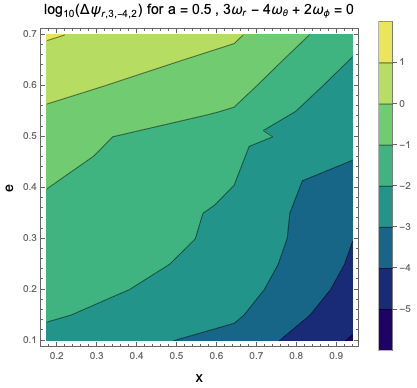}
\hskip 0.5cm
\includegraphics[width=0.3\linewidth]{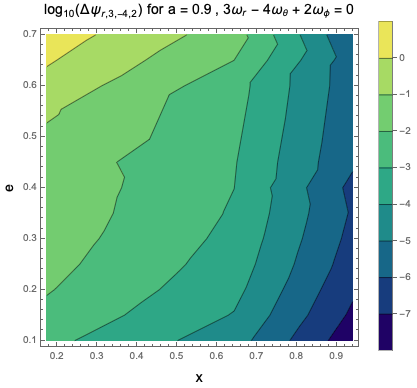}
\caption{Accumulated phase $\Delta \Psi_{i}$ for spin parameter $a = 0.1, 0.5, 0.9$ for a prograde orbit crossing the $3:-4:2$ resonance in the $x-e$ plane. Top, middle and bottom panels correspond to $\Delta \Psi_{\phi}$, $\Delta \Psi_{\theta}$ and, $\Delta \Psi_{r}$, respectively. The phase shift is computed for an EMRI with $M=4 \times 10^6 M_\odot, \mu =30 M_\odot$ under the influence of a tidal perturber with mass $M_\star=30 M_\odot$ at a distance of $10 \rm AU$ from the central SMBH. Results for different sets of parameters can be estimated from the scaling relation given in Eq.~\eqref{eq:scale}.}
\label{fig:PSpro342}
\end{figure*}

\begin{figure*}
 \centering
  \includegraphics[width=0.3\linewidth]{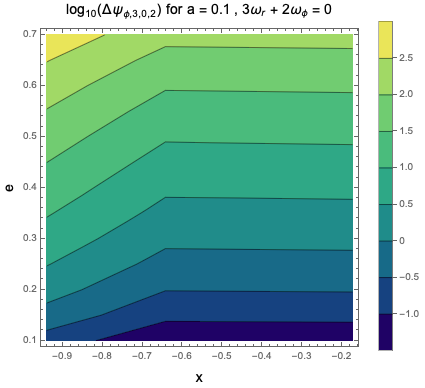}
    \hskip 0.5cm
\includegraphics[width=0.3\linewidth]{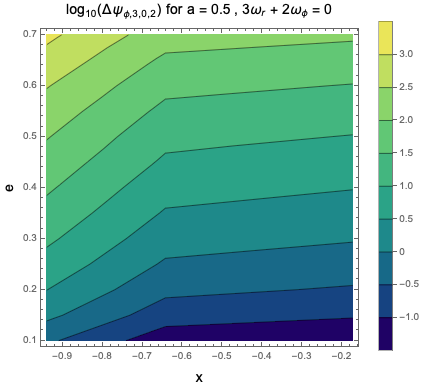}
\hskip 0.5cm
\includegraphics[width=0.3\linewidth]{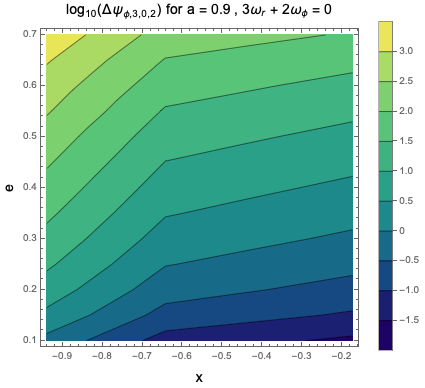}

\includegraphics[width=0.3\linewidth]{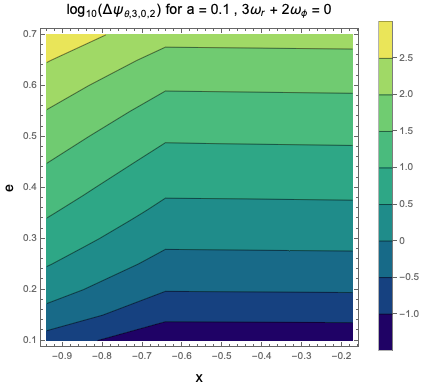}
    \hskip 0.5cm
\includegraphics[width=0.3\linewidth]{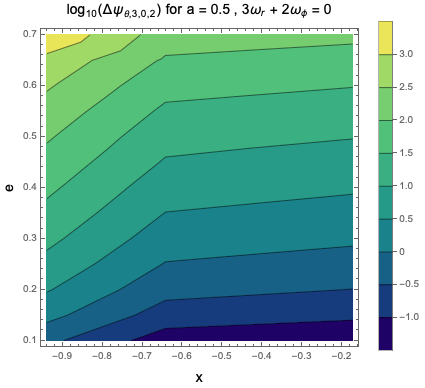}
\hskip 0.5cm
\includegraphics[width=0.3\linewidth]{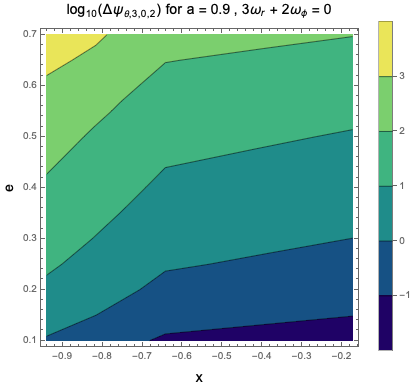}

\includegraphics[width=0.3\linewidth]{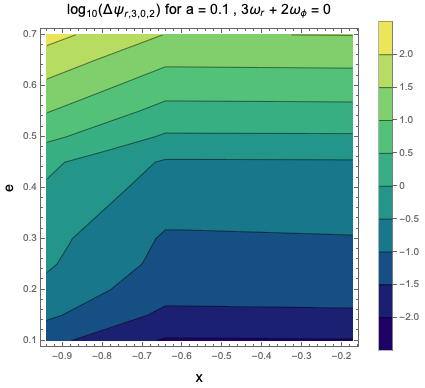}
    \hskip 0.5cm
\includegraphics[width=0.3\linewidth]{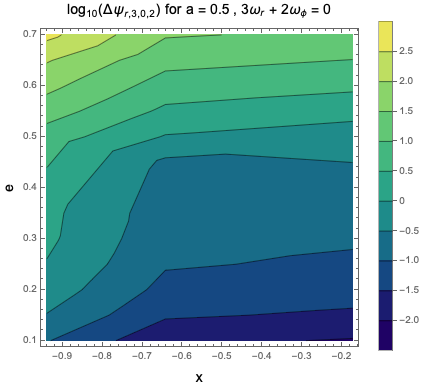}
\hskip 0.5cm
\includegraphics[width=0.3\linewidth]{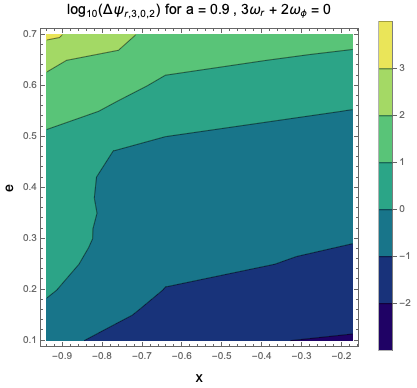}
\caption{Accumulated phase $\Delta \Psi_{i}$ for spin parameter $a = 0.1, 0.5, 0.9$ for a retrograde orbit crossing the $3:0:2$ resonance in the $x$ - $e$ plane. Top, middle and bottom panels correspond to $\Delta \Psi_{\phi}$, $\Delta \Psi_{\theta}$ and, $\Delta \Psi_{r}$, respectively. The phase shift is computed for an EMRI with $M=4 \times 10^6 M_\odot, \mu =30 M_\odot$ under the influence of a tidal perturber with mass $M_\star=30 M_\odot$ at a distance of $10 \rm AU$ from the central SMBH. Results for different set of parameters can be estimated from scaling relation given in Eq.~\eqref{eq:scale}.}
\label{fig:PSret302}
\end{figure*}

\begin{figure*}
  \centering
  \includegraphics[width=0.3\linewidth]{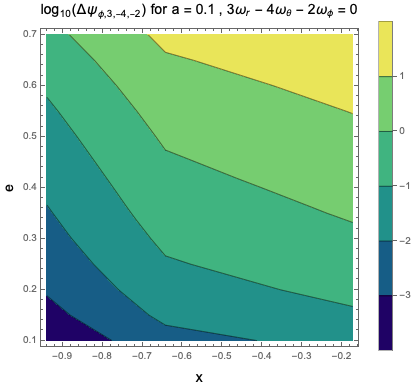}
    \hskip 0.5cm
\includegraphics[width=0.3\linewidth]{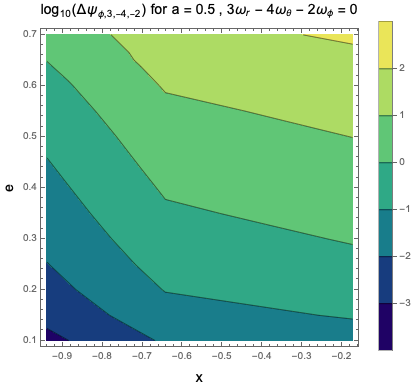}
\hskip 0.5cm
\includegraphics[width=0.3\linewidth]{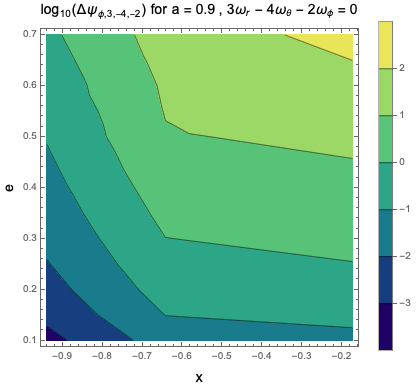}

\includegraphics[width=0.3\linewidth]{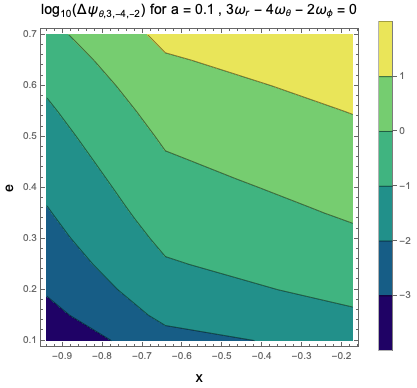}
    \hskip 0.5cm
\includegraphics[width=0.3\linewidth]{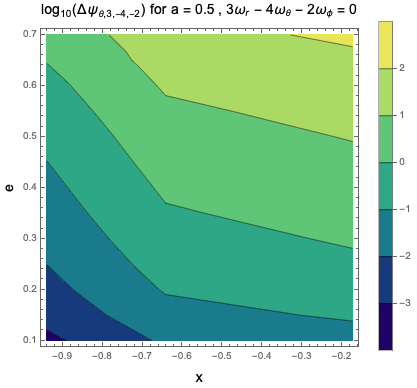}
\hskip 0.5cm
\includegraphics[width=0.3\linewidth]{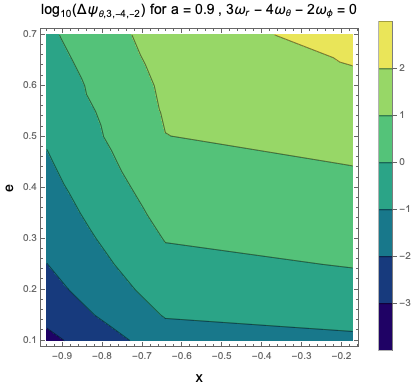}

\includegraphics[width=0.3\linewidth]{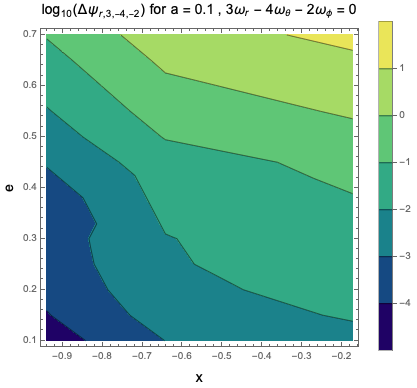}
    \hskip 0.5cm
\includegraphics[width=0.3\linewidth]{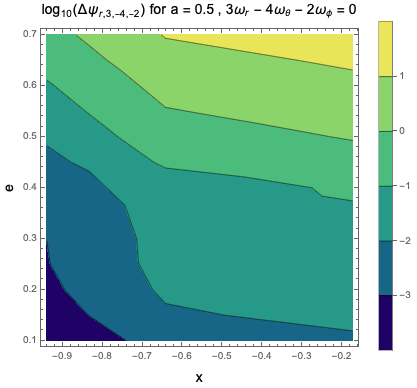}
\hskip 0.5cm
\includegraphics[width=0.3\linewidth]{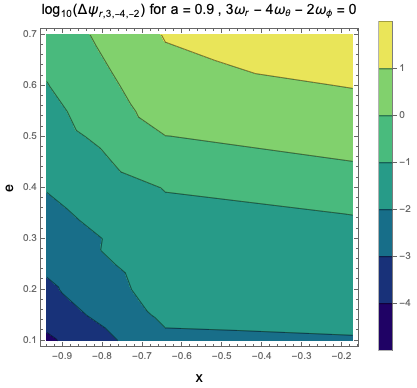}
\caption{Accumulated phase $\Delta \Psi_{i}$ for spin parameter $a = 0.1, 0.5, 0.9$ for a retrograde orbit crossing the $3:-4:-2$ resonance in the $x-e$ plane. Top, middle and bottom panels correspond to $\Delta \Psi_{\phi}$, $\Delta \Psi_{\theta}$ and, $\Delta \Psi_{r}$, respectively. The phase shift is computed for an EMRI with $M=4 \times 10^6 M_\odot, \mu =30 M_\odot$ under the influence of a tidal perturber with mass $M_\star=30 M_\odot$ at a distance of $10 \rm AU$ from the central SMBH. Results for different set of parameters can be estimated from scaling relation given in Eq.~\eqref{eq:scale}.}
\label{fig:PSret342}
\end{figure*}
For an EMRI source to be detectable by space-based interferometers, it must have an orbital frequency higher than about $f_{\rm LISA}=10^{-4} \rm Hz$. Using the approximation of Keplerian frequency when EMRI enters LISA band, we arrive at a rough condition on the semi-major axis $a_{\rm semi}$,
\beq
\frac{a_{\rm semi}}{M} < 20\times \bigg(\frac{M}{4 \times 10^6 M_\odot}\bigg)^{\!\!-2/3}  \bigg(\frac{f_{\rm LISA}}{10^{-4} \rm Hz}\bigg)^{\!\!-2/3}.
\eeq
 Using this rough estimate, an EMRI with $a_{\rm semi}$ less than $20M$ will lie in the observable band. Low-order resonances encountered by both prograde and retrograde orbits lie well within LISA frequency band for the central black hole less massive than $4\times 10^6 M_\odot$.

As discussed in previous sections, an orbit passing through a resonance can lead to a sudden change in constants of motion. This change means that the evolution post-resonance can become out of phase with that of the pre-resonance evolution. Therefore, we cannot match both parts with the same template. This can hamper the detection of EMRIs using standard matched filtering techniques. Thus, it is important to study their impact on EMRI waveforms. To estimate the effect, we study the deviation in the orbital phase, which can be evaluated as
\beq
\Delta \Psi_{\phi} = \int_{0}^{T_{\rm plunge}} 2 \Delta \omega_{\phi} dt\,.
\label{eq:phase}
\eeq

The accumulation in phase is integrated from the resonance time up to the plunge time $T_{\rm plunge}$. We evolve two orbits one with and without $\Delta J_{i}$ included. At each time $\omega_{\phi}$ for both the orbits is compared and the difference in frequencies for these two evolutions is given by $\Delta \omega_{\phi}$. The factor of 2 in Eq.~\eqref{eq:phase} is because the strongest harmonic in GWs is the quadrupolar mode ($l=2,m=2$). The phase evolution of waveform depends on the combination of three orbital phases: radial, polar, azimuthal. Therefore, in a similar manner, we also evaluate radial and polar accumulated phase shift, {\it i.e.}, $\Delta \Psi_{r}$ and $\Delta \Psi_{\theta}$, respectively. LISA has a remarkable sensitivity to the phase resolution of EMRI measurements, which is  roughly estimated as $\Delta \Psi_{\phi}\sim 0.1$, assuming SNR to be 20~\cite{Gair_2017,byh}. The resonance causes a shift in fundamental frequencies that is not replicated by adiabatic evolution, thus resulting in gradual dephasing of waveforms. 

In our analysis, we show that in a significant fraction of the parameter space EMRIs are likely to experience a tidal resonance (or multiple) that induces phase shift greater than $0.1$ rad making the effect detectable. Therefore, including the signature of resonances in waveform modeling is necessary to test GR with precision and allows a study of the environment around an EMRI. To compute the phase shift we set $M = 4 \times 10^6 M_\odot$, $\mu=M_\star=30 M_\odot$ and $R=10 \rm{AU}$. This distance as twice as far as in \cite{byh} to give a more conservative estimate. In Fig~\ref{fig:PSpro302}, the accumulation in phase is shown for prograde orbits crossing the $3:0:-2$ resonance in the $x$ - $e$ plane for different spin parameters of the SMBH. In the top panel, $\Delta \Psi_{\phi}$ is shown. The whole parameter space except for low eccentricity orbits ($< 0.2$) is affected by this resonance as the phase shift lies in the detectable range of LISA. Middle and bottom panel shows the affected parameter space for $\Delta \Psi_{\theta}$ and $\Delta \Psi_{r}$, respectively. The dephasing increases with increasing eccentricity and mildly depends on the spin parameter. Since this resonance is encountered early in the inspiral phase (see upper panel of Fig~\ref{fig:TidalReso}), the phase is accumulated over hundreds of thousands of cycles before plunge and therefore affects most of the parameter range. 

In Fig.~\ref{fig:PSpro342}, a similar plot is shown for a prograde orbit crossing the $3:-4:2$ resonance. In this case, dephasing is sensitive to changes in inclination and spin parameter. For the case $\Delta \Psi_{\phi}$ (top panel), orbits with low eccentricity ($\lsim 0.3$) and small inclination ($\lsim 45^\circ$) have phase shift smaller than 0.1, implying that the tidal resonance does not cause an observable effect in this range. As the spin is increased, a larger region of the parameter space is in the non-observable range. For $a=0.9$, only orbits with high inclination  ($\gsim 50^\circ$) and high eccentricity have a detectable tidal effect. The middle panel shows $\Delta \Psi_{\theta}$ which is of the same order as $\Delta \Psi_{\phi}$, and the bottom panel shows $\Delta \Psi_{r}$.

In Fig.~\ref{fig:PSret302} and Fig.~\ref{fig:PSret342}, we show the accumulated phase shift for retrograde orbits crossing the $3:0:2$ and $3:-4:-2$ resonances, respectively, for different spin parameters. As is clear from the figures, dephasing is larger compared to prograde orbits. This is expected because the value of $p$ is larger for retrograde orbits (see lower panel of Fig.~\ref{fig:TidalReso}), causing the effect of tidal force to be larger compared to prograde orbits. In contrast to the trend observed for prograde orbits, dephasing increases as the spin parameter increases.

The accumulated phase shown for different resonances in Figs.~\ref{fig:PSpro302}-\ref{fig:PSret342} is calculated for fixed masses of the SMBH, EMRI and the tidal perturber. The results can be translated for other masses using simple scaling. The change in phase is caused by the induced jump (see Eq.~\eqref{eq:Jump}) at resonance which scales as $\epsilon/\eta^{1/2}$. To compute accumulation in phase, we need to integrate over $1/\eta$ inspiral cycles. Therefore, the accumulated phase for a different set of parameters $\{M',\mu',M'_\star,R'\}$ is
\begin{align}
    \Delta \Psi'_{nkm} = \Delta \Psi_{nkm} \bigg(\frac{M'}{M}\bigg)^{\!\!7/2} \bigg(\frac{\mu'}{\mu}\bigg)^{\!\!-3/2} \bigg(\frac{M'_\star}{M_\star}\bigg) \bigg(\frac{R'}{R}\bigg)^{\!\!-3}.
\label{eq:scale}
\end{align}
Our results suggest that dephasing due to low-order tidal resonances should be easily detectable assuming that such tidal perturbers exist. The traditional adiabatic template will lose track of the phase evolution thereby lowering the signal-to-noise ratio after an EMRI encounters a resonance. We have shown the accumulation in phase shift for only one encounter of a tidal resonance, but, a realistic inspiral can undergo multiple resonances before plunge, further dephasing the signal. Thus, careful modeling of waveforms is needed to test GR with EMRI signals. In addition, such resonances can shed light on the stellar-mass distribution around galactic centers.

\section{Discussion}
\label{sec:5}
In the presence of a tidal perturber, an EMRI can encounter multiple resonances before plunge. Each resonance lasts for hundreds or thousands of orbital cycles depending on the EMRI's mass ratio. The effect of resonances (self-force and tidal) on phase evolution contributes more than post-adiabatic corrections. In this paper, we assessed the impact of tidal resonances on gravitational waves with the aim of surveying the orbital parameter space and investigating how often tidal resonances occur in realistic inspirals. We showed the dependence of resonances on the orbital phase and also found some trends such as the effect of spin of the central massive black hole, and the orbital parameters of the EMRI on the number of resonances encountered and the strength of each resonance. These trends are:
\begin{itemize}
  \item The resonance jump increases as the orbital eccentricity increases.
  \item As the orbital inclination angle increases the change in $Q$ and  $L_z$ increases for both prograde and retrograde orbits.
  \item For prograde orbits, as the spin parameter of the SMBH increases, the change in $L_{z}$ and $Q$ decreases. The opposite is true for retrograde orbits.
  \item Resonances with odd $k$ integers are suppressed and hence do not modulate the EMRI evolution.
\end{itemize}
Using these results, we computed the accumulation in phase after a tidal resonance has been encountered by an EMRI to understand their impact on waveforms. The study of dephasing revealed that less eccentric systems do not leave a detectable imprint in the phase evolution. We also provide fitting formulae for the change in the constants of motion caused by two low-order tidal resonances (see Eqs.~\eqref{eq:FittingLz302Pro}-\eqref{eq:FittingQ302Pro} and \cite{BHPC}), which can be efficiently used to take into account the resonance jump in waveform modeling without much computational cost. In addition to the semi-analytic calculations of the resonance jump, we have implemented the effect of the tidal perturber numerically using the forced osculating orbital elements method. This confirms that the tidal perturber only affects the EMRI significantly during resonances and agrees with the semi-analytic calculations of the jump size across a resonance.

This work is a first step towards understanding the observational importance of tidal resonances. We plan to extend this work by relaxing the assumption of a tidal perturber restricted to the equatorial plane, and by considering multiple resonant interactions with the same perturber at different points in time. While the forced osculating orbital elements method described in Sec. \ref{sec:num} is primarily used here to validate our analytical calculations (due to its higher computational cost), it is being used in ongoing work to explore various strategies and implications for waveform modeling and data analysis in the presence of a tidal perturber~\cite{PBAT}. In that work, we will characterize more fully the impact of tidal resonances on the search and inference for the EMRI itself (instead of merely focusing on the accumulated dephasing). We will also investigate the measurability of the tidal perturber's parameters, and devise optimal strategies for including tidal resonances in practical waveform models (the latter of which will be relevant for self-force resonance modeling as well). Based on the results in~\cite{speri2021assessing}, generic resonance jumps can be at least weakly constrained from EMRI observations, and so we are optimistic that suitable waveform models may allow $M_{\star}/R^3$ and the sky location of the tidal perturber to be measured in the case of stronger signals.

The Mathematica notebooks used for calculations and fitting formulae are available upon request.
\begin{acknowledgments}
We thank Ryuichi Fujita for sharing numerical data of GW fluxes. We are also grateful to Jonathan Gair, Niels Warburton, Philip Lynch and Soichiro Isoyama for sharing relevant code and for helpful discussions, as well as Huan Yang and Scott Hughes for feedback on our draft. This work makes use of the Black Hole Perturbation Toolkit~\cite{BHPToolkit}. PG is supported by MEXT scholarship. AJKC acknowledges support from the NASA grant 18-LPS18-0027. TT is supported by JSPS KAKENHI Grant Number JP17H06358 (and also JP17H06357), \textit{A01: Testing gravity theories using gravitational waves}, as a part of the innovative research area, ``Gravitational wave physics and astronomy: Genesis'', and also by JP20K03928. 
\end{acknowledgments}

\appendix
\section{Suppression of odd $k$ integer resonances}
\label{appex:A}

\begin{widetext}
\begin{figure*}
  \centering
  \includegraphics[width=0.3\linewidth]{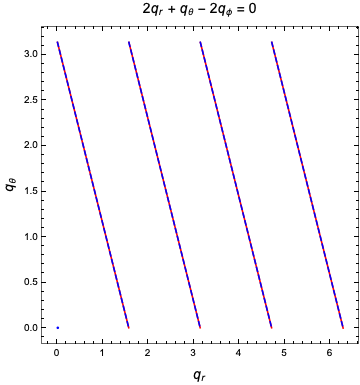}
    \hskip 0.5cm
\includegraphics[width=0.3\linewidth]{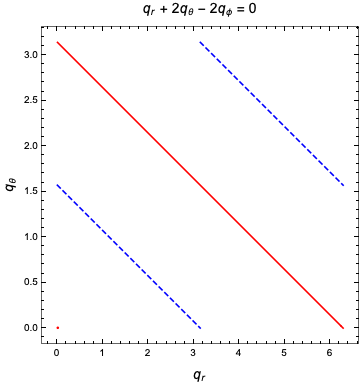}
\hskip 0.5cm
\includegraphics[width=0.3\linewidth]{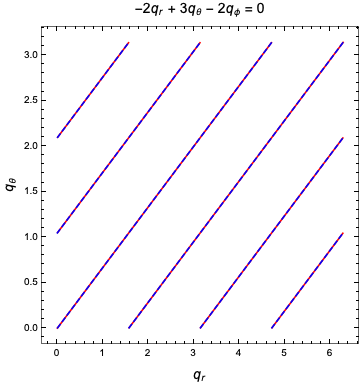}
\caption{Section of orbit in $q_r$ - $q_\theta$ plane for different resonance conditions. The red lines and blue dashed lines are obtained for $q_\phi=0$ and $q_\phi=\pi/2$, respectively.}
\label{fig:kodd}
\end{figure*}

\end{widetext}

We found that tidal resonances with odd $k$ integers do not give rise to a jump in the constants of motion. Hence, they do not contribute to a secular accumulation of a phase shift and are therefore not relevant for waveform modeling. In Fig.~\ref{fig:kodd}, for illustrative purpose, we show section of orbit in $q_{r}$ - $q_{\theta}$ plane for different resonances. In the leftmost panel we consider a $2:1:-2$ resonance (odd $k$) and compare section for fixed values of $q_{\phi}=0$ (red lines) and  $q_{\phi}=\pi/2$ (blue-dashed lines). On rotation of the orbit by $\pi/2$, the plot shows the same value for $q_r$ and $q_\theta$. Thus, the net tidal force of $m=\pm 2$ modes acting on the orbit cancels out completely resulting in no change in $L_z$. While this discussion is helpful in understanding the vanishing $d{L}_z/dt$ on crossing odd $k$ resonances, empirically we found that $d{Q}/dt$ also vanishes for such resonances.
The middle plot shows a $k=2$ resonance. In this case, two lines are not identical: therefore, the tidal force couples with the quadrupole moment of the orbit causing a finite jump in $L_z$. The rightmost plot shows the $-2:3:-2$ resonance exhibiting the same behavior as the $k=1$ case.

\nocite{*}
\newpage
\bibliography{ref}

\end{document}